\documentclass[%
preprint,
 amsmath,amssymb,
]{revtex4}

\usepackage{graphicx}
\graphicspath{{images/}}

\usepackage{mathrsfs}

\begin{document}

\title{Quantum proper time: A Finsler space from entropy and purity}

\author{Joseph Balsells\footnote{New address: Cornell Laboratory for
    Accelerator-based Sciences and Education,   Cornell University, Ithaca, New York 14853, USA}}
\email{jab934@cornell.edu}
\author{Martin Bojowald}
\email{bojowald@psu.edu}
\affiliation{%
  Institute for Gravitation and the Cosmos,
  The Pennsylvania State University, \\
  104 Davey Lab, 251 Pollock Road,
  University Park, Pennsylvania 16802, USA
}

\begin{abstract}
  A quantum clock cannot be modeled as a point mass moving along a single
  geodesic if it is in a state with nonzero position fluctuations. Instead, it
  is an extended object subject to tidal forces and a superposition of time
  dilations at different altitudes. Here, a geometrical formulation of quantum
  mechanics is used to show that additional quantum properties representing
  correlations between different directions imply a non-Riemannian geometrical
  structure experienced by a quantum clock. A specific version of Finsler
  geometry parametrized by entropy and purity of the state provides a novel
  setting for a combination of quantum and gravitational effects. A crucial
  ingredient is given by a new parametrization of quantum-information
  properties related to second-order moments of a state and may also be useful
  in other applications.
\end{abstract}


\maketitle

\section{Introduction}
\label{sec:introduction}

Geodesic motion of a quantum system in a background spacetime provides a tractable setting in which to investigate the interplay between quantum mechanics, gravitation, and quantum information. Compared with full quantum gravity, this problem remains sufficiently structured to permit detailed analysis while still capturing key conceptual tensions that any consistent quantum description of spacetime must address. From the perspective of quantum information, as we show in this work, this setting reveals that entropy, purity, and entanglement can be incorporated into a canonical geometric framework alongside classical phase-space variables. These quantities are influenced by relativistic and gravitational effects and, in turn, contribute to an effective description of motion. We show that this interplay is governed by an underlying Poisson structure in which entropy and purity arise as Casimir invariants, placing them within the same geometric framework as classical position and momentum.

The departure point for our description is the general relativistic notion of a test particle specified by its phase-space coordinates $(x^a, p_a)$. This system's motion between timelike events follows from the geometric principle of proper time. General relativity is sensitive only to the classical state of motion and measures the proper time interval as the square root of a position-dependent quadratic form of the classical tangent vectors,
\begin{equation}
\label{classicalProperTime}
\mathrm{d}\tau = \sqrt{-c^{-2} g_{ab}(x)\mathrm{d}x^a \mathrm{d}x^b}\,.
\end{equation}
This quadratic structure defines a pseudo-Riemannian geometry on spacetime. A natural question is whether general relativity---or at least the geometric principle of proper time---may be generalized to describe the motion of quantum systems.

The answer initially appears to be in the negative. A quantum system is not characterized by a single phase-space point. Instead, fluctuations contribute to the energy expectation value and therefore act as an effective gravitational source. Such effects have been incorporated in previous approaches, for example through path-integral formulations in which propagation is expressed as a sum over trajectories weighted around classical geodesics~\cite{HistoriesRelPart,PathIntPart,DynTheoryCurved}, and through semiclassical (WKB) expansions in which the phase defines classical geodesics while amplitude corrections are transported along them~\cite{SpinMotion}.

These approaches reduce quantum dynamics to families of classical trajectories, each equipped with a classical proper time, rather than introducing a single intrinsic geometric object replacing proper time for quantum systems. It remains unclear whether a framework exists in which quantum properties directly enter the geometry governing motion, generalizing Eq.~(\ref{classicalProperTime}). A genuinely geometric formulation of quantum dynamics may help identify the structural requirements of a quantum theory of gravity.

In a recent paper \cite{QuantumProperTime}, we initiated a program aimed at developing such a geometric framework for quantum proper time. There we showed that to first order in a semiclassical expansion quantum proper time can be obtained by extending classical Riemannian geometry to include degrees of freedom associated with quantum fluctuations. This extension replaces the classical metric $g_{ab}$ with a quasiclassical metric $g_{\bar{a}\bar{b}}$ that amends the classical coefficients by $\hbar$-terms and introduces new dimensions for fluctuation variables. The resulting quasiclassical metric determines not only the motion of expectation values in spacetime, but also the simultaneous evolution of the underlying quantum state.

For simplicity, our initial analysis was restricted to motion in one spatial dimension in a curved spacetime, illustrated by the explicit example of radial infall toward a black hole. In that setting, we obtained a specific quantum contribution to proper time as a function of fluctuation parameters, which in the Minkowski limit agreed qualitatively with results derived from wave-function-based approaches \cite{ClockSystemInt,ClassQuantProperTime}.

When additional spatial dimensions are included, however, new quantum features arise, such as correlations between spatial coordinates. These correlations modify both the trajectory of motion, through deviations from classical geodesics, and the accumulated proper time. Viewing the quantum state as an extended distribution on classical phase space, these effects can be interpreted as tidal perturbations that deflect the centroid from classical geodesic motion. Alternatively, we show here that the dynamics of a quantum state with two spatial degrees of freedom can be formulated as a point trajectory governed by a variational principle for quantum proper time in an extended phase space.

A key result is that this formulation requires not merely an extension of the geometric structure over new quantum degrees of freedom as in the previously studied 1-degree-of-freedom case but generically requires a genuine generalization from Riemannian to Finslerian geometry.
Unlike Riemannian geometry, which is defined by a metric tensor \(g_{ab}\) and a distance functional quadratic in the tangent vectors, Eq.~(\ref{classicalProperTime}), a Finsler geometry is characterized by a generalized distance functional for which \({\rm d}\tau^2\) is not necessarily quadratic in
\({\rm d}x^a\). Such geometries are known to give rise to modified dispersion relations and nontrivial light-cone structures; see, for example, Refs.~\cite{FinslerDispersion,FinslerGeometryDispersion}.

The paper is organized as follows. In Sec.~\ref{sec:setup} we review the Hamiltonian formulation of the Riemannian geodesic problem together with the geometrical formulation of relativistic quantum mechanics. In Sec.~\ref{sec:geom-quant-phase} we apply this quantization procedure to the Hamiltonian geodesic problem, first for a single degree of freedom and then for 2 degrees of freedom. In each case we introduce canonical parameters encoding the state-dependent modifications to proper time and inertial response of a quantum clock in a form that allows the kinetic dependence to be separated. The Finsler Hamiltonian governing the 2-degree-of-freedom system appears in this section as Eq.~(\ref{eq:quarticFinslerHamiltoniana}).

In Sec.~\ref{sec:inform-prop} we relate the canonical parametrization of the 2-degree-of-freedom system to symplectic quantum-optical operations. These considerations help us to show that the geodesic motion of a pure 2-degree-of-freedom Gaussian state is governed by an effectively Riemannian geometry that arises on the boundary of the space of quantum states. In contrast, generic mixed or non-Gaussian states require a genuinely Finslerian description. This result provides a new interpretation of purity conservation in Hamiltonian dynamics: pure states appear as boundary states, and Riemannian geometry emerges as a limiting case of the more general Finsler geometry.
In Sec.~\ref{sec:time-dilation} we return to the main problem and present quantum-corrected gravitational time-dilation formulas, including their Finsler-geometric interpretation.

% We identify expressions for entropy, purity, and various correlations in a quasiclassical treatment, relating these physical concepts to specific functions of second-order moments of a state for two independent classical degrees of freedom. While the expressions for entropy and purity are known from corresponding formulas for Gaussian states, we obtain a new correlation parameter, related to different kinds of uncertainty products, that is distinguished by its role in the Finsler nature of our extended geometry.

\section{Setup}
\label{sec:setup}

We begin with a brief review of relevant ingredients from
\cite{QuantumProperTime}, to which we refer for more details.

\subsection{Hamiltonian formulation}

The classical functional of proper time $\tau$ for a given worldline $\gamma$
parametrized by $\sigma$ in the range $[0,1]$ and in spacetime with a Riemannian
metric $g_{ab}$ is given by
\begin{equation}
  \label{eq:classicalProperTime}
  \tau[\gamma] 
  = 
  \int_0^1 \sqrt{-\frac{1}{c^2}g_{ab}(x) \frac{{\rm d}x^a}{{\rm d}\sigma} \frac{{\rm
        d}x^b}{{\rm d}\sigma}} {\rm d}\sigma\,.
\end{equation}
The velocity term can be replaced by momenta if the Hamiltonian 
\begin{equation}
  \label{eq:classicalProperTimeHamiltonFunction}
  H(x^a,p_a)
  = -\frac{1}{2m}
  \left(
    g^{ab}p_a p_b + (mc)^2
  \right)
\end{equation}
is introduced, where
\begin{equation} \label{pa}
  p_a= -mg_{ab}\dot{x}^b
\end{equation}
with a derivative by proper time $\tau$ indicated by the dot.  Hamilton's
equations, $\dot{x}^a=\partial H/\partial p_a$ and
$\dot{p}_a=-\partial H/\partial x^a$, can then be seen to be equivalent to the
second-order geodesic equation if (\ref{pa}) is used to replace $p_a$ in the equation
for $\dot{p}_a$. Equation (\ref{pa}) follows independently from the
equation for $\dot{x}^a$.

The original integral (\ref{eq:classicalProperTime}) has a symmetry because
the wordline $\gamma$ can be parametrized by any parameter $\sigma$,
which may be redefined using a monotonic function
$\sigma'=f(\sigma)$. As a consequence, the Hamiltonian
(\ref{eq:classicalProperTimeHamiltonFunction}) not only generates equations of
motion, it also has to vanish for allowed solutions $(x^a,p_b)$ and therefore
implies a Hamiltonian constraint, $H(x^a,p^a)=0$. In a region of weak
gravitational field, where
$g_{ab}\approx\eta_{ab}$ is close to the Minkowski metric, the constraint then
implies the mass-shell condition 
\begin{equation} \label{p0}
  \frac{E^2}{c^2}=|\vec{p}|^2+m^2c^2
\end{equation}
for the energy $E=cp^0=-cp_0$ with the spatial part $\vec{p}$ of the 4-momentum $p_a$.

Given the Hamiltonian constraint, an inverse Legrendre transformation implies
the Lagrangian
\begin{equation}\label{L}
  L= - \frac{1}{2} mg_{ab}\dot{x}^a\dot{x}^b  + \frac{1}{2}mc^2\,.
\end{equation}
The proper-time functional can therefore be expressed as
\begin{equation} \label{tau2}
  \tau[\gamma]= \int \sqrt{ \frac{2L}{mc^2} - 1 }\;{\rm d}\tau\,.
\end{equation}
This expression can be interpreted in two ways. (i) If we insert the Lagrangian
(\ref{L}), we rewrite the integral (\ref{tau2}) in the form
(\ref{eq:classicalProperTime}). (ii) If the constraint $H=0$ is used, it
implies $L= mc^2$, such that (\ref{tau2}) equals $\int {\rm d}\tau$ and indeed
represents proper time.

Our strategy is now to compute quantum corrections implied by a canonical
quantization of the Hamiltonian $H$, and derive implications for the
proper-time functional after an inverse Legendre transformation. We will
maintain the constraint condition that the Hamiltonian vanish, making sure
that we still obtain a reparametrizable theory for an expression that
measures proper time. However, using an explicit expression for the quantum
Hamiltonian, the combination $g_{ab}\dot{x}^a\dot{x}^b$ will be amended by
quantum corrections.

These corrections can be computed systematically in a
quasiclassical formulation \cite{EffAc,Karpacz,Bosonize,EffPotRealize} in which the expectation value
$\langle\hat{H}\rangle$ in a generic semiclassical state is parametrized by
central position-momentum moments up to a fixed order, given here by second
order. Endowing moments with a canonical structure then allows us to identify
momenta of quantum variables such as quantum fluctuations. Using lessons
from the one-dimensional treatment in \cite{QuantumProperTime}, we expect to
arrive at an expression of the form
\begin{equation}
  2m\langle\hat{H}\rangle + m^2c^2 = -g^{\bar{a}\bar{b}}_{\mathcal{Q}}p_{\bar{a}}p_{\bar{b}}
\end{equation}
for the proper-time integrand. Here, the bar indicates that additional degrees
of freedom from quantum parameters appear, and the subscript $\mathcal{Q}$
indicates $\hbar$-corrections in the metric coefficients. Perturbatively in
$\hbar$, the extended metric remains invertible and can be used for a
derivation of the Lagrangian and the velocity dependence of proper time. As a
surprising new result, we will demonstrate that the quantum theory introduces an additional quartic term in $\langle\hat{H}\rangle$ of
the form
$g^{\bar{a}\bar{b}\bar{c}\bar{d}}p_{\bar{a}}p_{\bar{b}}p_{\bar{c}}p_{\bar{d}}$. This term is the source of non-Riemannian
behavior in the Finsler framework.

\subsection{Quantum theory}
\label{sec:quantum-theory}

Since the classical Hamiltonian is constrained to vanish, we have to apply
constrained quantization. In practice, this means that only three of the
momentum components are turned into operators because the fourth one, like
$p_0$ in the mass-shell condition (\ref{p0}), is not independent; see
\cite{EffCons,EffConsRel,QuantumProperTime} for details. As a
consequence, quantum corrections depend on fluctuations, correlations, and
other parameters only for the three spatial momenta in $\vec{p}$, even in the
relativistic case. We derive these corrections using a geometrical
formulation of quantum mechanics that determines the state space as a Poisson
manifold parametrized by moments of a state.

Given a quantum state, which may be represented by a wave function or a
density matrix, we can compute the moments
\begin{eqnarray}
  \Delta \big (x^{\alpha}p^{\beta}\big)&=&\Delta \big ((x^1)^{\alpha_1}(x^2)^{\alpha_2}
                                           (x^3)^{\alpha_3}p_1^{\beta_1}p_2^{\beta_2}
                                           p_3^{\beta_3} \big)\nonumber \\
  &=& \big\langle
      (\hat{x}^1-x^1)^{\alpha_1}(\hat{x}^2-x^2)^{\alpha_2}(\hat{x}^3-x^3)^{\alpha_3}
      (\hat{p}_1-p_1)^{\beta_1}(\hat{p}_2-p_2)^{\beta_2}(\hat{p}_3-p_3)^{\beta_3} 
      \big\rangle_{\rm symm}.
      \label{eq:momentdef}
\end{eqnarray}
Here, the subscript \({\rm symm}\) indicates complete symmetrization of the expectation value and we use multi-index notation with
\(\alpha = (\alpha_1,\alpha_2, \alpha_3)\) and
\(\beta = (\beta_1,\beta_2,\beta_3)\), so that
\begin{equation}
  x^\alpha = (x^1)^{\alpha_1} (x^2)^{\alpha_2} (x^3)^{\alpha_3} \quad\mbox{and}\quad
  p^\beta =  (p_1)^{\beta_1} (p_2)^{\beta_2} (p_3)^{\beta_3}\,.
\end{equation}
Conversely, a pure or mixed state can in principle be reconstructed from a set
of moments together with basic expectation values,
$( \vec{x}, \vec{p}, \Delta(x^\alpha p^\beta))$ where $\vec{x}=\langle\hat{\vec{x}}\rangle$
and $\vec{p}=\langle\hat{\vec{p}}\rangle$, provided the latter fulfill uncertainty
relations and higher-order analogs. Basic expectation values and moments can
therefore be used as coordinates on state space.

We define a semiclassical state as one for which the moments exist and fulfill the
hierarchy
\begin{equation}
  \label{eq:hierarchy}
  \Delta(x^\alpha p^\beta) = O(\hbar^{\vert \alpha + \beta \vert /2})
\end{equation}
with the moment order
\begin{equation}
  \label{order}
  \vert \alpha \vert =
   \alpha_1 + \alpha_2 + \alpha_3
\end{equation}
of multi-indices, which can be added component by component. A quasiclassical
expansion up to a given order in $\hbar$ then allows us to truncate the set of
all moments to finite size. In what follows, we will work with first order in
$\hbar$, or second order in moments.

For a canonical formulation, we define a Poisson bracket on the space of moments as \cite{EffAc,Karpacz}
\begin{equation}\label{Poisson}
  \{\langle\hat{A}\rangle,\langle\hat{B}\rangle\}=\frac{\langle[\hat{A},\hat{B}]\rangle}{i\hbar}
\end{equation}
and extend it to products of expectation values by using the product
rule. This definition implies that quantum evolution of the form
\begin{equation}
  \frac{{\rm d}\langle\hat{O}\rangle}{{\rm
      d}\tau}=\frac{\langle[\hat{O},\hat{H}]\rangle}{i\hbar}
\end{equation}
is generated by the quantum Hamiltonian
$H_{\mathcal{Q}}=\langle\hat{H}\rangle$, parametrized as a function of
moments through its state dependence.

In practice, writing the expectation value of a Hamilton operator in a generic
state as a function of moments is a challenging, if not impossible, task. For
semiclassical states, however, the result can be obtained by a Taylor
expansion in
\begin{eqnarray}
  H_\mathcal{Q}\big (x^a,p_a,\Delta(x^\alpha p^\beta) \big )
  &=& \left\langle
    H \big (x^a + (\hat{x}^a - x^a),p_a + (\hat{p}_a - p_a)\big )\right\rangle
    \nonumber \\[1em] 
  &=& \sum_{\alpha=0}^\infty\sum_{\beta=0}^\infty
    \frac{1}{\alpha!\beta !} \frac{\partial^{\vert \alpha+\beta \vert }H(x,p)}{\partial^{\alpha}x\partial^{\beta}p} \; \Delta(x^\alpha p^\beta) \label{eq:Hq}
\end{eqnarray}
where we assume that the Hamilton operator is ordered completely
symmetrically and use the multi-index definition
\begin{equation}
  \label{factorial}
  \alpha! = \alpha_1! \alpha_2 ! \alpha_3!\,.
\end{equation}
This Hamiltonian generates coupled evolution equations for basic expectation values as
well as moments,
\begin{equation}
  \frac{{\rm d} \Delta(x^{\alpha}p^{\beta})}{{\rm d}\tau} =
  \{\Delta(x^{\alpha}p^{\beta}),H_{\mathcal{Q}}\}\,. 
\end{equation}

\section{Geometry of the quantum phase space}
\label{sec:geom-quant-phase}

The quantum geodesic Hamiltonian computed using (\ref{eq:Hq}) and (\ref{eq:classicalProperTimeHamiltonFunction}) with the lowest-order nonvanishing quantum corrections takes the form
\begin{eqnarray}
  \label{eq:Hq2dof}
    &&H_\mathcal{Q}\big (t,\vec{x},p_0,\vec{p}, \Delta(x^ix^j), \Delta(x^ip_j),
  \Delta(p_ip_j) \big )\nonumber\\ 
    &\approx & -\frac{1}{2m}
        \left(
        g^{ab}p_ap_b + m^2c^2
        +g^{ij}\Delta(p_ip_j)
        +\frac{\partial g^{aj}}{\partial x^k} p_a\Delta(x^kp_j)
        + \frac{1}{2} \frac{\partial^2g^{ab}}{\partial x^k\partial x^l}p_ap_b \Delta(x^kx^l)
        \right) \,.
\end{eqnarray}
This expression includes sums over four-vector indices \(a\) and \(b\) as well as
over spatial indices \(i,j,k\), and \(l\). [A strict constrained quantization
following \cite{EffConsRel} would rather amount to expanding the Hamiltonian
expression obtained for $p_0$ after solving the constraint $H=0$, which as a
square root would imply additional coefficients for the moment terms compared
with (\ref{eq:Hq2dof}). However, the coefficients are well approximated by our
expression, and easier to obtain, provided the spatial momentum squared,
$g^{ij}p_ip_j$, is much smaller than $m^2c^2$, which is what we assume in what
follows. These coefficients are relevant in the massless limit, but they do
not affect properties of the moments where quantum-information properties
appear, constituting the main interest of this paper.]

From a geometrical perspective, the central question is whether the expression
in (\ref{eq:Hq2dof}) is in some sense quadratic in momenta. If this is the
case, it can be written as $(2m)^{-1}(g^{ab}_{\mathcal Q}p_ap_b+m^2c^2)$ with
some metric $g^{ab}_{\mathcal Q}$ extending the metric of the underlying
classical theory. The structure of spacetime would then remain Riemannian in
the presence of quantum corrections of a test mass.

However, while each term in the expression except for $m^2c^2$ contains two
symbols of $p$, it is not immediately clear whether momentum (co)variances
$\Delta(p_ip_j)$ can be considered quadratic in a new momentum, or whether the
covariances $\Delta(x^kp_j)$ are linear in some quantum momentum.
Nevertheless, it is possible to address this question in an unambiguous manner
because the moment space is a phase space equipped with a Poisson bracket
derived from (\ref{Poisson}).

A momentum component is defined as a phase-space coordinate \(p_b\), which is
part of a momentum vector that satisfies the canonical Poisson bracket
relation \(\{x^a, p_b\} = \delta_b^a\) with the position vector
\(x^a\). The relevance of having such pairs of coordinates is that they imply equations of
motion in Hamilton's form, where the time derivative of one coordinate is
given by plus or minus the partial derivative of the Hamiltonian by the
conjugate variable in the pair. The appearance of partial derivatives then
implies a notion of independence of different pairs. 

\subsection{Spacetime geometry for single-mode states}
\label{sec:quant-single-class}

The phase-space variables \(\Delta \big (x^{\alpha}p^{\beta}\big)\) are not canonical coordinates because they have nontrivial Poisson brackets with one another. For instance, the \(x\)-coordinates at second order, \((\Delta(x^2),\Delta(xp_x),\Delta(p_x^2))\), have the brackets
\begin{equation} \label{eq:oneDoFBrackets}
  \{\Delta(x^2),\Delta(p_x^2)\}=4\Delta(xp_x)\,,\quad
  \{\Delta(x^2),\Delta(xp_x)\}=2\Delta(x^2)\,,\quad
  \{\Delta(xp_x),\Delta(p_x^2)\}=2\Delta(p_x^2).
\end{equation}

In this one-dimensional restriction, it is known \cite{VariationalEffAc,GaussianDyn,QHDTunneling} that the transformation
\begin{equation}
  \label{eq:canonicalCoordinates}
  \Delta \left(x^2\right) = s^2
  \qquad
  \Delta (x p_x) = s p_s
  \qquad
  \Delta \left(p_x^2\right) = p_s^2 + \frac{U}{s^2}
\end{equation}
to new variables \((s,p_s,U)\) implies the canonical Poisson bracket $\{s,p_s\}=1$ making \((s,p_s)\) a canonical pair. The third variable, $U$, Poisson commutes with all second-order moments, making \(U\) an element of the center of the Poisson algebra generated by (\ref{eq:oneDoFBrackets}). 
In a Poisson algebra, central elements are known as Casimir functions,
in analogy with the corresponding concept in Lie algebras. Casimir functions
represent conserved quantities as they are invariant under any Hamiltonian flow. These Casimir functions can be systematically identified for any algebra.

The Poisson algebra, given by Eqs.~(\ref{eq:oneDoFBrackets}), is isomorphic to the
Lie algebra \(\mathrm{sp}(2,\mathbb{R})\), providing a natural connection to the quantum-optical formalism.
In this setting, a single bosonic mode is described in terms of quadrature operators $\hat{x}$ and $\hat{p}_x$, which satisfy the canonical commutation relations.  The associated single-mode covariance matrix
\begin{equation}
  \label{eq:33}
  \sigma = \left(
    \begin{array}{cc}
      \Delta(x^2) & \Delta(xp_x) \\
      \Delta(xp_x) & \Delta(p_x^2)
    \end{array}
    \right)
\end{equation}
together with the symplectic form \(\Omega_{ij} = \{\xi_i,\xi_j\}\) [with
\({\bf \xi} = (x,p_x)\) having components $\xi_i$] allow the Casimir of
\(\mathrm{sp}(2,\mathbb{R})\) to be constructed equivalently either as the sum
of squared eigenvalues of \(\Omega \sigma\) or as
\(\mathrm{tr}\left[(\Omega \sigma)^2\right]\).
%The eigenvalues of $\Omega \sigma$ are referred to as the symplectic eigenvalues of $\sigma$.
More generally, the algebra \(\mathrm{sp}(2n,\mathbb{R})\) has \(n\) Casimirs that are constructed as
\begin{equation}
  \label{eq:generalspRcasimirs}
  U^{(2k)} \propto \mathrm{Tr} \left[(\Omega\sigma)^{2k}\right], \quad k=1,\ldots, n.
\end{equation}
By construction, these quantities are invariant under symplectic
transformations.

Inverting the mapping (\ref{eq:canonicalCoordinates})
identifies the single-mode Casimir as the uncertainty product
\begin{equation}
  U=\Delta(x^2)\Delta(p_x^2)-\Delta(xp_x)^2
\end{equation}
which is, by Eq.~(\ref{eq:generalspRcasimirs}), a quadratic Casimir.
% In the two-mode case the quadratic Casimir includes additional contributions from the second mode and additionally a second, quartic Casimir appears.
As a direct consequence of the uncertainty relation, the single-mode Casimir is bounded from
below by $\hbar^2/4$. We may therefore write
\begin{equation} \label{pq}
  U = p_q^2 \frac{\hbar^2}{4}
\end{equation}
with a momentum $p_q\geq 1$, canonically conjugate to a coordinate $q$ that
does not appear in the Hamiltonian. [It is convenient to separate $p_q$ and
$\hbar$ in (\ref{pq}) such that $\hbar$ can be used to characterize
semiclassical orders. As a consequence, both \(p_q\) and \(q\) are unitless.]

If we consider only quantized motion along the $x$-direction (or the radial direction in a
spherical coordinate system), we can apply the one-dimensional canonical mapping (\ref{eq:oneDoFBrackets}) to express the geodesic quantum Hamiltonian in terms of momenta as
\begin{eqnarray}
    &&H_\mathcal{Q}\big (t,\vec{x},p_0,\vec{p}, \Delta(x^2), \Delta(xp_x),
  \Delta(p_x^2) \big )\nonumber\\ 
    &=& -\frac{1}{2m} \left( g^{ab}p_ap_b + m^2c^2
      +g^{xx}\Delta(p_x^2) + 2\frac{\partial g^{ax}}{\partial
       x} p_a\Delta(xp_x)+ \frac{1}{2} \frac{\partial^2g^{ab}}{\partial x^2}p_ap_b
        \Delta(x^2)\right)\nonumber\\
    &=& -\frac{1}{2m} \left( g^{ab}p_ap_b + m^2c^2
      +g^{xx}\left(p_s^2+\frac{\hbar^2}{4s^2}p_q^2\right) + 2\frac{\partial g^{ax}}{\partial
       x} sp_ap_s+ \frac{1}{2} \frac{\partial^2g^{ab}}{\partial x^2}s^2p_ap_b\right) \,. \label{eq:Hq1dof}
\end{eqnarray}
This expression is strictly quadratic in momentum components, which now
include $p_s$ and $p_q$. These quantum corrections can therefore be expressed
by an extended six-dimensional metric 
\begin{equation} \label{eq:gQinv}
  g_{\mathcal{Q}}^{\bar{a}\bar{b}} :=
  \begin{pmatrix}
    g^{tt} + \frac{1}{2}s^2 \partial_x^2g^{tt}
    & g^{tx} + \frac{1}{2}s^2 \partial_x^2g^{tx} & g^{ty} + \frac{1}{2}s^2 \partial_x^2g^{ty}
    & g^{tz} + \frac{1}{2}s^2 \partial_x^2g^{tz} & s \partial_x g^{tx} & 0
    \\
    g^{tx} + \frac{1}{2}s^2 \partial_x^2g^{tx}
    & g^{xx} + \frac{1}{2}s^2 \partial_x^2 g^{xx} & g^{yx} + \frac{1}{2}s^2 \partial_x^2g^{yx}
    & g^{zx} + \frac{1}{2}s^2 \partial_x^2 g^{zx} &s\partial_x g^{xx} & 0
    \\
    g^{ty} + \frac{1}{2}s^2 \partial_x^2g^{ty}
    & g^{xy} + \frac{1}{2}s^2 \partial_x^2 g^{xy} & g^{yy} + \frac{1}{2}s^2 \partial_x^2g^{yy}
    & g^{zy} + \frac{1}{2}s^2 \partial_x^2 g^{zy} &s\partial_x g^{yx} & 0 
    \\
    g^{tz} + \frac{1}{2}s^2 \partial_x^2g^{tz}
    & g^{xz} + \frac{1}{2}s^2 \partial_x^2 g^{xz} & g^{yz} + \frac{1}{2}s^2 \partial_x^2g^{yz}
    & g^{zz} + \frac{1}{2}s^2 \partial_x^2 g^{zz} &s\partial_x g^{zx} & 0
       \\
    s \partial_x g^{tx} & s\partial_x g^{xx} &s \partial_x g^{yx} & s\partial_x g^{zx} & g^{xx} & 0
    \\ 
    0&0&0 & 0 & 0 & \frac{1}{4}\hbar^2 g^{xx}/s^2
  \end{pmatrix}.
\end{equation}
such that
\begin{equation}
  \label{eq:HQgeodesic}
  H_{\mathcal{Q}}(t,x,y,z,s,q,p_t,p_x,p_y,p_z,p_s,p_q) = -\frac{1}{2m} \left(
    g_{\mathcal{Q}}^{\bar{a}\bar{b}} p_{\bar{a}}p_{\bar{b}} + m^2c^2\right) 
\end{equation}
with an extended $p_{\bar{a}} = (p_t,p_x,p_y,p_z,p_s,p_q)$.

The upper-left block of the pseudo-Riemannian metric is obtained by applying a truncated Taylor expansion operator to the classical metric coefficients,
\begin{equation}
  \label{eq:9}
  \langle g^{ab}(\hat{x}) \rangle
  =
  \left\langle
    \exp \left[
      \left( \hat{x} - \langle \hat{x} \rangle \right) \frac{{\rm d}}{{\rm d}\hat{x}}
    \right]
  \right\rangle
  g^{ab}(\langle\hat{x}\rangle)
  = \left(1 + \frac{1}{2} \Delta(x^2) \partial_x^2\right) g^{ab}(x) + O(\hbar^{3/2}).
\end{equation}
This series expansion is directly analogous to the moment expansion of $H(\hat{x})$ in Eq.~(\ref{eq:Hq}). With this identification, the effective inverse metric takes the form
\begin{equation} \label{eq:gQinvCondensed}
  g_{\mathcal{Q}}^{\bar{a}\bar{b}} =
  \begin{pmatrix}
    \langle g^{ab}(\hat{x}) \rangle  &  s\partial_x g^{ax} & 0    \\
    s\partial_x g^{xb} & g^{xx} & 0
    \\ 
    0& 0 & \frac{1}{4}\hbar^2 g^{xx}/s^2
  \end{pmatrix}.
\end{equation}

This expression defines an effective metric \(g_{\mathcal{Q}}^{\bar{a}\bar{b}}\) governing the free motion of a quantum test particle on a curved classical background. The upper-left block is the expectation value of the classical inverse metric, evaluated componentwise in the particle's quantum state. As a result, rather than interacting directly with the classical background, the quantum particle experiences a smoothed, state-averaged geometry. Furthermore, unlike a classical point particle, a quantum object carries additional degrees of freedom. Specifically, Eq.~(\ref{eq:gQinvCondensed}) allows for a single spatial fluctuation coordinate \(s\) and its corresponding uncertainty coordinate \(q\).

The extension of the metric over nonclassical degrees of
freedom has several implications. First, the path-length functional gives quantum perturbations to proper time
\begin{equation}
  \label{eq:HamiltonianProperTime}
  \tau_{\mathcal{Q}} =
  \int \sqrt{1 + \frac{2H_{\mathcal{Q}}}{mc^2}} {\rm d}\tau.
\end{equation}
The form of \(H_{\mathcal{Q}}\) from Eq.~(\ref{eq:HQgeodesic}) together with
the effective quantum inverse metric (\ref{eq:gQinvCondensed}) was used in
Ref.~\cite{QuantumProperTime} to compute corrections to gravitational time
dilation that might be detectable by near-term optical atomic clocks.

As a further consequence, quantum fluctuations in the Hamiltonian constraint \(H_{\mathcal{Q}} = 0\) lead to the modified dispersion relation
\begin{equation} \label{eq:quantumDispersion}
  \left( g^{ab}
    + \frac{1}{2}  \Delta(x^kx^l) \frac{\partial^2g^{ab}}{\partial x^k \partial x^l}
  \right)  p_ap_b
  + g^{ij}\Delta(p_ip_j)
  + 2\frac{\partial g^{aj}}{\partial x^k} p_a\Delta(x^kp_j)
   + m^2c^2
  = 0.
\end{equation}

In a region of weak gravitational field, where $g_{ab}\approx\eta_{ab}$ is
close to the Minkowski metric, corrections containing the metric derivative
drop out and the mass-shell condition becomes
\begin{equation} \label{eq:quantumDispersionWeak}
p_0^2 = |\vec{p}|^2+ \delta^{ij}\Delta(p_ip_j) + m^2c^2.
\end{equation}
This result captures the contribution of quantum momentum fluctuations to the energy so
that even classically stationary particles have nonzero ``zero-point'' energy.
The fluctuation term can be interpreted as contributing to an effective mass
\begin{equation}
  \label{eq:EffectiveMass}
  m_{\text{eff}}^2c^2 = m^2c^2 + \delta^{ij}\Delta(p_ip_j),
\end{equation}
in which case even classically massless objects (light or a gravitational
wave) acquire some effective mass from nonvanishing momentum fluctuations. (As
already mentioned, precise coefficients for a massless case require a
systematic implementation of the formalism given in \cite{EffConsRel}.)

From an experimental point of view, distinguishing quantum uncertainty contributions in Eq.~(\ref{eq:quantumDispersionWeak}) from classical line-broadening has so far been challenging. In the nonrelativistic limit, Eq.~(\ref{eq:quantumDispersionWeak}) expands as
\begin{equation} \label{eq:quantumDispersionWeakNonRelativistic}
  E \approx mc^2
  + \frac{|\vec{p}|^2}{2m}
  + \frac{\delta^{ij}\Delta(p_ip_j)}{2m}
  + \frac{|\vec{p}|^2\delta^{ij}\Delta(p_ip_j) }{4m^3c^2}.
  %   - \frac{|\vec{p}|^4}{8m^3c^2}
\end{equation}
The zeroth-order momentum term acts as a mass correction, as in (\ref{eq:EffectiveMass}). The expression Eq.~(\ref{eq:quantumDispersionWeakNonRelativistic}) can be compared with the model-independent generalized dispersion relation,
\begin{equation}
  \label{eq:30}
  E = mc^2 + \frac{\vert \vec{p}\vert^2}{2m} + \frac{1}{2M_P}
  \left(
    \xi_1 m\vert \vec{p}\vert c + \xi_2\vert \vec{p}\vert^2
  \right)
\end{equation}
where \(M_P\) is the Planck mass, and \(\xi_1,\xi_2\) are dimensionless
parameters introduced in Ref.~\cite{DispersionConstraints}.
The parameter \(\xi_2\) constrains the quadratic momentum-dependent term.
Cold-atom recoil experiments currently provide
95\% confidence bounds of \(-6.0 < \xi_1 < 2.4\) and \(\vert \xi_2 \vert <
10^9\). The limit on \(\xi_2\) implies
\begin{equation}
  \label{eq:15}
  \frac{M_P}{m} \frac{\delta^{ij}\Delta(p_ip_j)}{2m^2c^2} = \xi_2 < 10^9,
\end{equation}
which, in the specific case of cesium atoms (\(M_P/m_{\rm Cs} \approx 10^{17}\)), translates to the bound
\begin{equation}
  \label{eq:15b}
  \Delta(p_x^2) + \Delta(p_y^2) + \Delta(p_z^2)  < 300~\mathrm{MeV}^2/c^2.
\end{equation}
This bound is comparable to the momentum spread of a thermal distribution at room temperature,
\begin{equation}
  \label{eq:doppler-broadening}
  \sigma_p^2 = m k_{\rm B} T
\end{equation}
for \(T \approx 300~\mathrm{K}\). Employing cooling or velocity-selective saturation spectroscopy techniques, the classical linewidth can be reduced well below this threshold.

Atom interferometry experiments measuring photon recoil on atomic states prepared in large-momentum superpositions could yield more stringent bounds on nonclassical corrections to dispersion relations than the limit established in Ref.~\cite{DispersionConstraints}. Such experiments would involve cooling atoms, applying a beam-splitting \(\pi/2\) pulse coupled to an appropriate electronic transition to generate momentum-space superpositions, and subsequently measuring the atomic energy distribution. The required atomic temperature depends directly on the achievable magnitude of the momentum splitting, which must ensure that classical Doppler broadening described by Eq.~(\ref{eq:doppler-broadening}) remains smaller than the quantum-uncertainty-induced broadening. Experimental confirmation of the weak-field quantum contributions expressed in Eq.~(\ref{eq:quantumDispersionWeak}) would constitute evidence that nonclassical degrees of freedom inherent to quantum states experience a modified and extended spacetime geometry.

\subsection{Spacetime geometry for two-mode states}
\label{sec:quant-two-class}

Our identification of the effective inverse metric
(\ref{eq:gQinvCondensed}) demonstrated that the effective spacetime quantum theory for localized single-mode quantum
states preserves the pseudo-Riemannian structure of the classical theory at the lowest semiclassical order. As shown in \cite{Bosonize,EffPotRealize}, the
quadratic dependence on momenta persists when extending moments of a single classical degree of freedom $(x,p)$ up to at least fourth order, such as \(\Delta(x^4)\). However, according to the same papers, this property no
longer holds when two independent classical degrees of freedom, $(x,p_x)$ and $(y,p_y)$, are quantized. In contrast to single-mode systems, bipartite systems inherently involve entanglement entropy, whose presence, as we demonstrate below, requires a departure from the quadratic structures underlying Riemannian geometry.

\subsubsection{Canonical structure of second-order statistics}
\label{sec:canon-struct-second}

In a quantum system with two continuous degrees of freedom, there are ten second-order moments: six describing the individual modes and four capturing cross-mode correlations
\begin{eqnarray}
  \label{eq:momentVector}
  \Delta_i&=&\big(
  \Delta(x^2),
  \Delta(xp_{x}),
  \Delta(p_{x}^2),
  \Delta(y^2),
  \Delta(yp_{y}),
      \Delta(p_{y}^2), \\
  &&\quad
  \Delta(xy),
  \Delta(xp_{y}),
  \Delta(p_{x}y),
  \Delta(p_{x}p_{y})
\big). \nonumber
\end{eqnarray}
% These degrees of freedom are naturally organized into the two-mode covariance matrix,
% \begin{equation}
%   \label{eq:sigma}
%   \sigma =
%   \begin{pmatrix}
%     \sigma_x & \sigma_{xy} \\
%     \sigma_{xy}^\top & \sigma_y
%   \end{pmatrix}.
% \end{equation}
% Here \(\sigma_x\) and \(\sigma_y\) are the single-mode covariance matrices for each degree of freedom, while \(\sigma_{xy}\) encodes their correlations:
% \begin{equation}
%   \label{eq:7}
%   \sigma_x =
%   \begin{pmatrix}
%     \Delta(x^2) & \Delta(xp_x) \\
%     \Delta(x p_x) & \Delta(p_x^2)
%   \end{pmatrix}, \quad
%   \sigma_y =
%   \begin{pmatrix}
%     \Delta(y^2) & \Delta(yp_y) \\
%     \Delta(y p_y) & \Delta(p_y^2)
%   \end{pmatrix}, \quad
%   \sigma_{xy} =
%   \begin{pmatrix}
%     \Delta(xy) & \Delta(xp_y) \\
%     \Delta(p_x y) & \Delta(p_xp_y)
%   \end{pmatrix}.
% \end{equation}
% The Poisson tensor, \(\mathbb{P}_{ij} = \{ \Delta_i, \Delta_j\}\), has rank eight, indicating that a canonical mapping will identify from the ten moments \(\Delta_i\) four canonical pairs: \((s_x,p_{s_x})\), \((s_y,p_{s_y})\), \((\alpha,p_\alpha)\), \(( \beta, p_\beta)\). In a faithful
% mapping of the ten second-order moments, there additionally must be two Casimir functions, \(C_1\) and \(C_2\), labelling the symplectic leaves.
Under the Poisson bracket in Eq.~(\ref{Poisson}) the ten second-order moments form a Lie-Poisson manifold isomorphic to $\mathrm{sp}(4,\mathbb{R})^*$.
Since the Poisson tensor \(\mathbb{P}_{ij} = \{ \Delta_i, \Delta_j\}\) has rank eight a generic symplectic leaf is eight-dimensional and is labeled by two independent Casimirs, denoted here by $C_1$ and $C_2$. References~\cite{Bosonize,EffPotRealize} introduced the canonical variables
$(s_x,p_{s_x})$, $(s_y,p_{s_y})$, $(\alpha,p_\alpha)$, and $(\beta,p_\beta)$ to
provide a Darboux chart on such a leaf. This chart is adapted to distinguished embedded
subalgebras of $\mathrm{sp}(4,\mathbb{R})$, namely two one-mode
$\mathrm{sp}(2,\mathbb{R})$ sectors, a compact $\mathrm{su}(2)$ passive
mode-mixing sector, and a noncompact $\mathrm{su}(1,1)$ two-mode squeezing
sector. 

The first two pairs $(s_x, p_{s_x})$ and $(s_y, p_{s_y})$, generating two copies of $\mathrm{sp}(2,\mathbb{R})$, are defined analogously to the single-degree-of-freedom mapping as
\begin{equation} \label{eq:2dofMapping1}
  \Delta(x^2) = s_x^2
  \,,\qquad
  \Delta (y^2) = s_y^2
  \,,\qquad
  \Delta(xp_x) = s_x p_{s_x}
  \,,\qquad
  \Delta(yp_y) = s_y p_{s_y}\,,
\end{equation}
so that $p_{s_x}$ and $p_{s_y}$ are the one-mode squeezing momenta conjugate to the widths.

Intermode correlations introduce additional degrees of freedom beyond two single-mode sectors.
First, the position covariance $\Delta(xy)$ is parametrized by a canonical variable $\beta$ according to
\begin{equation}
  \Delta(xy) = s_x s_y \cos(\beta)\,.
\end{equation}
The off-diagonal mixed moments then decompose into a piece inherited from the one-mode structure and a genuinely two-mode contribution,
\begin{equation}
  \Delta(xp_y) = s_x p_{s_y} \cos(\beta)
  - \sin(\beta)\frac{s_x}{s_y}\left(p_{\alpha}+p_{\beta}\right),
\end{equation}
\begin{equation}
  \Delta(p_xy) = p_{s_x}s_y \cos(\beta)
  + \sin(\beta)\frac{s_y}{s_x}\left(p_{\alpha}-p_{\beta}\right).
\end{equation}
The intermode generators $p_\beta$ and $p_\alpha$ enter these terms with definite parity:
$p_\beta$ appears symmetrically in both expressions and therefore generates correlated pair excitations characteristic of the noncompact $\mathrm{su}(1,1)$ sector associated with two-mode squeezing, whereas $p_\alpha$ appears with opposite signs and generates number-conserving rotations between the modes in the compact $\mathrm{su}(2)$ sector.

The momentum-momentum correlations constitute the genuinely constrained part of this construction. Their form is fixed by the symplectic structure once the above parametrizations are imposed. Explicitly, they are
\begin{equation}
  \label{eq:Deltapx2}
  \Delta(p_x^2)=p_{s_x}^2+ \frac{(p_{\alpha}-p_{\beta})^2}{s_x^2}
  +\frac{1}{2 s_x^2 \sin^2{(\beta)}}
  \left(
    C_1^2-4p_{\alpha}^2
    - \sqrt{P}\sin{(\alpha+\beta)}
  \right),
\end{equation}

\begin{equation}
  \label{eq:Deltapy2}
  \Delta(p_y^2)=p_{s_y}^2
  + \frac{(p_{\alpha}+p_{\beta})^2 }{s_y^2} 
  + \frac{1}{2s_y^2\sin^2(\beta)}
  \left(
    C_1^2-4 p_{\alpha}^2
    - \sqrt{P}\sin(\alpha-\beta)
  \right),
\end{equation}
and
\begin{eqnarray}
  \label{eq:Deltap1p2}
  \Delta(p_x p_y)&=&\left(p_{s_x}p_{s_y} 
                     +\frac{p_{\alpha}^2-p_{\beta}^2}{s_xs_y}\right)\cos(\beta)
                 +\left(
                     \frac{p_{s_y}}{s_x}(p_{\alpha}-p_{\beta})
                     - \frac{p_{s_x}}{s_y}(p_{\beta}+p_{\alpha})\right)\sin(\beta)\nonumber\\
               &&-\frac{1}{2s_x s_y\sin^2(\beta)}\left((C_1^2-4p_{\alpha}^2)\cos(\beta) -
                  \sqrt{P} \sin(\alpha)\right).
\end{eqnarray}
These are the only moments to depend explicitly on the rotation parameter $\alpha$. Each of these moments involves the square root of a quartic momentum polynomial,
\begin{equation}
  \label{eq:canonicalSquareRoot}
  \sqrt{P(p_\alpha,C_1,C_2)} = \sqrt{(C_1^2 - 4p_{\alpha}^2)^2- (C_1^4-C_2^4)}\,.
\end{equation}
Special cases under which the polynomial \(P(p_\alpha,C_1,C_2)\) is factorizable are examined in Sec.~\ref{sec:riem-spec-cases}.

The above construction can be understood as a canonical realization of the Bloch-Messiah decomposition, in which a general symplectic transformation is factorized into passive rotations and active squeezing operations. 
In the present coordinates, $(s_x,p_{s_x})$ and $(s_y,p_{s_y})$ describe local squeezing, $(\beta,p_\beta)$ encodes nonlocal correlations, and $(\alpha,p_\alpha)$ parametrizes the residual passive mixing between the modes.

\subsubsection{The two-mode spacetime Finsler Hamiltonian}
\label{sec:two-mode-spacetime}

The canonical mapping
Eqs.~(\ref{eq:2dofMapping1})--(\ref{eq:Deltap1p2}) enlarges the cotangent space from the four-dimensional space spanned by the momentum covector \(p_{a} =(p_t,p_x,p_y,p_z)\) to a ten-dimensional space  spanned by the extended momentum covector
\begin{equation}
  \label{eq:32}
  p_{\bar{a}} =(p_t,p_x,p_y,p_z,p_{s_x},p_{s_y},p_\alpha, p_\beta, C_1, C_2),
\end{equation}
with the Casimir variables viewed formally as momenta conjugate to cyclic coordinates.

Substituting this mapping into the quantum geodesic
Hamiltonian, Eq.~(\ref{eq:Hq2dof}), yields a function that is not quadratic in the momenta. Instead, the presence of the square root \(\sqrt{P}\) leads to a generalized quartic structure~\cite{FinslerTextbook},
\begin{equation}
  \label{eq:quarticFinslerHamiltoniana}
  H_{\mathcal{Q}} = -\frac{1}{2m} \left (
    \sqrt{A^{\bar{a}\bar{b}\bar{c}\bar{d}}
      p_{\bar{a}}p_{\bar{b}}p_{\bar{c}}p_{\bar{d}}}
    + B^{\bar{a}\bar{b}} p_{\bar{a}}p_{\bar{b}}  + m^2 c^2 \right).
\end{equation}
Although this structure no longer induces a Riemannian geometry, its degree-two homogeneity in the momenta ensures that geometric
notions such as connection, parallel transport, and geodesic still remain well defined (see, e.g.,
\cite{FinslerGravity,FinslerGravityVariation}). % We provide explicit expressions for both \(A^{\bar{a}\bar{b}\bar{c}\bar{d}}\) and \(B^{\bar{a}\bar{b}}\) using the canonical mapping.

The fully symmetric covariant rank-four object
\(A^{\bar{a}\bar{b}\bar{c}\bar{d}}\) encodes the nonquadratic contribution to \(H_{\mathcal{Q}}\) and arises from the components of \(g^{ij}\Delta(p_ip_j)\) proportional to $\sqrt{P}$.
In terms of the kernel
\begin{equation}
  \label{eq:Kernel}
  K_{ij}(s_x,s_y,\alpha,\beta) = \csc^2(\beta )
  \left(
    \begin{array}{cc}
      -\sin (\alpha +\beta )/s_x^2 & \sin (\alpha ) /(s_x s_y) \\
      \sin (\alpha ) /(s_x s_y) & - \sin (\alpha -\beta )/s_y^2
    \end{array}
  \right),
\end{equation}
the quartic contribution to the Hamiltonian can be written compactly as
\begin{equation}
  \label{eq:quarticPart}
  \sqrt{A}
     = \frac{1}{2} K_{ij}g^{ij}\sqrt{P}\,.
\end{equation}
Its sign is determined by the contraction with the inverse metric,
\begin{equation}
  \label{purityWeightSign}
  \frac{1}{2} K_{ij}g^{ij} = -\frac{\sin (\alpha +\beta ) g^{xx}}{2 s_x^2}
  + \frac{\sin (\alpha ) g^{xy}}{s_x s_y}
  - \frac{\sin (\alpha -\beta ) g^{yy}}{2 s_y^2}.
\end{equation}
This term underlies the apparent ``weight of purity'' discussed in the companion paper \cite{EntropyPurity}.

The coefficients of the symmetric bilinear form \(B^{\bar{a}\bar{b}}\) may be arranged in the block matrix
\begin{equation}
  \label{eq:gQinv2dof}
  B^{\bar{a}\bar{b}} =
  \begin{pmatrix}
    \langle g^{ab}(\hat{x}, \hat{y}) \rangle
    & D[g^{aj}]
    &
      E
    & 0 &0 \\
    D[ g^{ib}] & \tilde{g}^{ij} & C & 0 & 0\\
    E^T
    & C^T & \Theta & 0 & 0 \\
    0 & 0 & 0 & \frac{1}{2} \mathrm{Tr}\left(\Sigma_x^{-1} g^{-1}\right) & 0 \\
    0 & 0 & 0 & 0 & 0
  \end{pmatrix}\,.
\end{equation}
In this expression \(a\) and \(b\) are spacetime indices taking
values \(t,x,y,z\), and \(i\) and \(j\) take values $x$ or $y$ indicating the quantized
mode. The explicit form of the block entries of (\ref{eq:gQinv2dof}) is given in Appendix \ref{a:block-entries-two}.

The matrix \(B^{\bar{a}\bar{b}}\) is noninvertible because $C_2$ appears only in the quartic part. Physically however, it is the fundamental inverse tensor
\begin{equation}
  \label{eq:25}
  g_{\mathcal{Q}}^{\bar{a}\bar{b}}(x,p)
  := m\frac{\partial^2 H_{\mathcal{Q}}}{\partial p_{\bar{a}} \partial p_{\bar{b}}}
\end{equation}
which determines meaningful geometric quantities. For the generalized quartic (\ref{eq:quarticFinslerHamiltoniana}),
\begin{equation}
  \label{eq:26}
  g_{\mathcal{Q}}^{\bar{a}\bar{b}}
  = B^{\bar{a}\bar{b}}
  + 3A^{\bar{a}\bar{b}} - 2A^{\bar{a}}A^{\bar{b}},
\end{equation}
where the normalized partial contractions \(A^{\bar{a}}\), \(A^{\bar{a}\bar{b}}\), and \(A^{\bar{a}\bar{b}\bar{c}}\) are defined through the relations
\begin{equation}
  \label{eq:Aa}
  A^{3/4} A^{\bar{a}} = A^{\bar{a}\bar{b}\bar{c}\bar{d}}  p_{\bar{b}}p_{\bar{c}}p_{\bar{d}}\,,
\end{equation}
\begin{equation}
  \label{eq:Aab}
  A^{1/2} A^{\bar{a}\bar{b}} = A^{\bar{a}\bar{b}\bar{c}\bar{d}} p_{\bar{c}}p_{\bar{d}}\,,
\end{equation}
\begin{equation}
  \label{eq:Aabc}
  A^{1/4} A^{\bar{a}\bar{b}\bar{c}} = A^{\bar{a}\bar{b}\bar{c}\bar{d}} p_{\bar{d}}\,,
\end{equation}
with the complete contraction
\begin{equation}
  \label{eq:A}
  A := A^{\bar{a}\bar{b}\bar{c}\bar{d}}  p_{\bar{a}}p_{\bar{b}}p_{\bar{c}}p_{\bar{d}}.
\end{equation}
The expression \(g_{\mathcal{Q}}^{\bar{a}\bar{b}}\) computed this way is invertible.

Similar to the one-variable expression, Eq.~(\ref{eq:gQinvCondensed}), the quadratic part of the generalized quartic Finsler metric for two variables depends explicitly on the correlation variables \(s_x\) and \(s_y\). However, the two-variable quadratic form (\ref{eq:gQinv2dof}) depends also on the spatial correlation angle \(\beta\), which introduces anisotropic behavior into the effective geometry. This anisotropy implies that the effective quantum geometry depends on the alignment (or misalignment) of \(x\) and \(y\) correlations.

The expression for the quadratic part of the effective quantum metric
simplifies when the correlation structure of the quantum state aligns
maximally with the slicing of the classical metric. In this case, \(\beta =
\pi /2\) [implying \(\Delta(xy) = 0\), an isotropic state], the quadratic part of the effective quantum metric is
\begin{eqnarray}
  \label{eq:14}
  &&B^{\bar{a}\bar{b}} =\\
  &&
  \left(
    \begin{array}{ccccccc}
      \langle g^{ab}(\hat{x}, \hat{y}) \rangle
      & s_x \partial_x  g^{ax}
      & s_y \partial_y  g^{ay}
      & \frac{s_y \partial_y g^{ax}}{s_x}-\frac{s_x \partial_x g^{ay}}{s_y}
      &  -\frac{s_y \partial_y g^{ax}}{s_x}-\frac{s_x \partial_x g^{ay}}{s_y}
      & 0 & 0 \\
      s_x \partial_x g^{xb}
      & g^{xx}
      & 0
      & -\frac{g^{xy}}{s_y}
      & -\frac{g^{xy}}{s_y}
      & 0 & 0 \\
      s_y \partial_y g^{yb}
      & 0
      & g^{yy}
      & \frac{g^{xy}}{s_x}
      & -\frac{g^{xy}}{s_x}
      & 0 & 0 \\
      \frac{s_y \partial_y g^{xb}}{s_x}-\frac{s_x \partial_x g^{yb}}{s_y}
      & -\frac{g^{xy}}{s_y}
      & \frac{g^{xy}}{s_x}
      & -\frac{g^{xx}}{s_x^2}
        -\frac{g^{yy}}{s_y^2}
      & \frac{g^{yy}}{s_y^2}-\frac{g^{xx}}{s_x^2}
      & 0 & 0 \\
      -\frac{s_y \partial_y g^{xb}}{s_x}-\frac{s_x \partial_x g^{yb}}{s_y}
      & -\frac{g^{xy}}{s_y}
      & -\frac{g^{xy}}{s_x}
      & \frac{g^{yy}}{s_y^2}-\frac{g^{xx}}{s_x^2}
      & \frac{g^{xx}}{s_x^2}+\frac{g^{yy}}{s_y^2}
      & 0 & 0 \\
      0 & 0 & 0 & 0 & 0
      & \frac{1}{2} \left( \frac{g^{xx}}{s_x^2}+\frac{g^{yy}}{s_y^2} \right) & 0 \\
      0 & 0 & 0 & 0 & 0 & 0 & 0 \\
    \end{array}
  \right)\,.\nonumber
\end{eqnarray}

\subsection{Finsler geometry}
\label{sec:implications}

In the physics literature, non-Riemannian Finsler metrics have previously been
considered in situations where spacetime geometry exhibits higher-order
dependencies on direction, motivated for instance by modified dispersion
relations \cite{FinslerDispersion,FinslerGeometryDispersion}. One example is the Randers metric
\cite{Randers}
\begin{equation}
  \label{eq:randers}
  \mathrm{d }s = k_a \mathrm{d }x^a + \sqrt{ \pm g_{ab}(x) \mathrm{d }x^a\mathrm{d }x^b},
\end{equation}
where the field \(k_a\) parametrizes small Lorentz violations. Particle
mechanics in the geometry (\ref{eq:randers}) obey a dispersion relation
different from the standard relativistic one. The Randers ansatz has been applied
in connection with the GZK threshold \cite{FinslerGZK} and
in cosmological studies \cite{FinslerFriedmann}.

Finsler metrics based on higher-order polynomials have also been studied. In
multimetric geometry \cite{FinslerMultiMetric} one encounters the Finsler structure
\begin{equation}
  \label{eq:11}
  F = \sum_\mu F^{(\mu)}
\end{equation}
in which each \(F^{(\mu)}\) is a Riemannian Finsler structure,
\(F^{(\mu)}(x,y) = \sqrt{g_{ij}^{(\mu)}(x) y^iy^j}\) where \((x^i,y^i)\) are
local coordinates on the tangent bundle of
spacetime. In the bimetric case with
\(F^{(1)} = \sqrt{g_{ij}^{(1)}(x) y^iy^j}\) and
\(F^{(2)} = \sqrt{g_{ij}^{(2)}(x) y^iy^j}\), the Finsler structure may be
rewritten
\begin{equation}
  \label{eq:BimetricFinsler}
  F(x,y) = \sqrt{\sqrt{4g_{ij}^{(1)}g_{kl}^{(2)}y^iy^jy^ky^l} +
    \left(g_{ij}^{(1)} + g_{ij}^{(2)} \right) y^iy^j}\,, 
\end{equation}
a special case of the generalized fourth-root Finsler structure
\begin{equation}
  \label{eq:genQuarticFinsler}
  F(x,y) = \sqrt{\sqrt{a_{ijkl}y^iy^jy^ky^l} + b_{ij} y^iy^j}
\end{equation}
studied in \cite{Finslerm,Finsler4}. The Finsler geometry found here is
precisely of this form, except that we have derived a Hamiltonian version. 

On a Finsler manifold, the natural Lagrangian structure \(L^F = F^2/2\) is
complemented by the dual Finsler structure \(F^*\), defined on the cotangent
bundle, which gives rise to the Hamiltonian \(H^F = (F^*)^2/2\). Finsler
Hamiltonians are discussed in detail in \cite{FinslerHamiltonian}. When
applied to (\ref{eq:genQuarticFinsler}), the result can be compared to our
Hamiltonian construction.

In our results, Finsler properties are not realized for the standard four
spacetime dimensions. They are instead controlled by quantum entropy and
purity through the conserved momenta $C_1$ and $C_2$, and experienced for
motion in the correlation dimension parametrized by $\alpha$. These
quantum-information properties will be discussed in the remainder of this paper.

\section{Canonical quantum information and geometric regimes of quantum state space}
\label{sec:inform-prop}

% Alternative titles :
% Canonical structure and geometric regimes of quantum state space
% Canonical quantum information in phase space

In this section, we connect our canonical parametrization of second-order moments to standard concepts in quantum information theory. Quantities such as entropy, purity, and entanglement admit natural expressions in terms of second moments, particularly for Gaussian states. Our mapping of second moments to canonical variables organizes these degrees of freedom into independent geometric sectors, $(s_x, p_{s_x}), (s_y, p_{s_y}) , (\alpha, p_\alpha), (\beta, p_\beta)$, together with the Casimirs $C_1$ and $C_2$. This structure provides a direct link between quantum information measures and the geometry of quantum state space.

\subsection{Reconstruction of canonical variables from second moments}
\label{sec:inverse-mapping}

The inverse mapping of Eqs.~(\ref{eq:2dofMapping1})--(\ref{eq:Deltap1p2}) includes the expressions
\begin{gather}
  \label{eq:MomentInverseA}
  s_x = \sqrt{\Delta(x^2)} \quad , \quad p_{s_x} = \frac{\Delta(xp_{x})}{\sqrt{\Delta(x^2)}} \\
  s_y = \sqrt{\Delta(y^2)} \quad , \quad p_{s_y} = \frac{\Delta(yp_{y})}{\sqrt{\Delta(y^2)}} \\
  \cos{(\beta)} = \frac{\Delta(xy)}{\sqrt{\Delta(x^2) \Delta(y^2)}} \\
  p_\alpha = \frac{\Delta(xy)(\Delta(yp_{y}) - \Delta(xp_{x})) + \Delta(x^2) \Delta(p_{x}y) - \Delta(y^2) \Delta(xp_{y})}
  {2\sqrt{\Delta(x^2) \Delta(y^2) - \Delta(xy)^2}}
  \label{eq:MomentInversepAlpha}\\
  p_\beta = \frac{\Delta(xy)(\Delta(yp_{y}) + \Delta(xp_{x})) - \Delta(x^2) \Delta(p_{x}y) - \Delta(y^2) \Delta(xp_{y})}
  {2\sqrt{\Delta(x^2) \Delta(y^2) - \Delta(xy)^2}}.
  \label{eq:MomentInverseB}
\end{gather}

The Casimirs $C_1$ and $C_2$ are conserved quantities and therefore functions of the symplectic eigenvalues $\nu_\pm$, defined as the magnitudes of the eigenvalues of \(i\Omega\sigma\) with $\sigma$ the two-mode covariance matrix. One finds
\begin{gather} \label{eq:quadraticCasimirC12}
  C_{1}^2 = \nu_+^2 + \nu_-^2 \\
  C_{2}^2 = \nu_+^2 - \nu_-^2.
  \label{eq:quarticCasimirC22}
\end{gather}
These are expressed explicitly as functions of the second-order moments by
\begin{equation}
  \label{eq:quadraticCasimir}
  \begin{aligned}
    C_1^2
    &= 
      \Delta(x^2) \Delta(p_x^2) - \Delta(xp_x)^2 \\
    &\qquad +
      \Delta(y^2) \Delta(p_y^2) - \Delta(yp_y)^2 \\
    &\qquad +
      2\Delta(xy)\Delta(p_xp_y)
      -
      2\Delta(xp_y)\Delta(p_xy) .
  \end{aligned}
\end{equation}
and
\begin{eqnarray}
  \label{eq:12}
  &&C_2^4 =
 \\
  && 4 \big(\Delta(p_xp_y) (\Delta(xp_x)-\Delta(yp_y))-\Delta(p_x^2) \Delta(xp_y)+ \Delta(p_y^2)\Delta(p_xy) \big)
\nonumber\\
  \enspace&&\times\big( \Delta(xy) (\Delta(yp_y)-\Delta(xp_x)) + \Delta(x^2) \Delta(p_xy) - \Delta(y^2)\Delta(xp_y)\big)
\nonumber\\
  \enspace&&+4 \big(\Delta(p_xy) (\Delta(xp_x)+\Delta(yp_y)) -\Delta(p_x^2) \Delta(xy)- \Delta(y^2)\Delta(p_xp_y)\big)
\nonumber\\
  \enspace&&\times\big(\Delta(xp_y)(\Delta(xp_x) + \Delta(yp_y)) - \Delta(p_xp_y) \Delta(x^2)-\Delta(p_y^2) \Delta(xy)\big)
     \nonumber\\
  \enspace&&+2 \left( \Delta(x^2)\Delta(p_x^2) - \Delta(xp_x)^2
             +\Delta(p_xp_y) \Delta(xy)-\Delta(p_xy) \Delta(xp_y)\right)^2
     \nonumber\\
  \enspace&&+2 \left(\Delta(p_xp_y) \Delta(xy)-\Delta(p_xy) \Delta(xp_y)
             +\Delta(y^2) \Delta(p_y^2) -\Delta(yp_y)^2\right)^2
     \nonumber\\
  \enspace&&+C_1^4\nonumber
\end{eqnarray}

The square of the first Casimir represents the state's full uncertainty and is bounded from below by $\hbar^2/2$ as a consequence of the two-mode uncertainty relation
\begin{equation}
  \label{RSuncertainty}
  \sigma + \frac{i \hbar}{2} \Omega \ge 0
\end{equation}
which implies that the symplectic eigenvalues satisfy \(\nu_+ \ge \nu_- \ge \hbar/2\).
% We may therefore
% write $C_1^2$ as
% \begin{equation} \label{pC1}
%   C_1^2 = p_{C_1}^2 \frac{\hbar^2}{2}
% \end{equation}
% with a momentum $p_{C_1}\geq 1$, canonically conjugate to a coordinate $q_1$ that
% does not appear in the Hamiltonian.
The second Casimir  measures the balance of uncertainty in each quadrature and is bounded below by 0. The lengthy expression for \(C_2\) is a consequence of its relation to the determinant of the covariance matrix according to
\begin{equation}
  \label{eq:detsigma}
  \det(\sigma) = \frac{1}{4}(C_1^4-C_2^4).
\end{equation}
For physical states, the covariance matrix must be positive, \(\sigma > 0\), which requires \(\det(\sigma)>0\). In fact, a stronger bound \(\det(\sigma)>\hbar^4/16\) holds~\cite{QuantumContinuous}. Consequently, \(C_1 = C_2\) is not physically realizable, with implications discussed in Sec.~\ref{sec:riem-spec-cases}.

To determine the dependence of $\alpha$ on the second moments we start from Eqs.~(\ref{eq:Deltapx2})--(\ref{eq:Deltap1p2}) for the momentum covariances $\Delta(p_ip_j)$ and isolate the terms containing trigonometric factors. Define the residuals
\begin{widetext}
  \begin{eqnarray}
    \label{eq:1}
    R_1 &:=&
             2 s_1^2 \sin^2(\beta)
             \left[
             \Delta(p_1^2) -
             p_{s_1}^2
             +
             \frac{(p_{\alpha}-p_{\beta})^2}{s_1^2}
             \right]
             -
             (C_1^2-4 p_{\alpha}^2)\,, \\
    R_2 &:=&
             2 s_2^2 \sin^2(\beta)
             \left[
             \Delta(p_2^2) -
             p_{s_2}^2
             +
             \frac{(p_{\alpha}+p_{\beta})^2}{s_2^2}
             \right]
             -
             (C_1^2-4 p_{\alpha}^2)\,, \\
    R_{12} &:=&
                2 s_1s_2 \sin^2(\beta)
                \Bigg[
                \Delta(p_1p_2) -
                \left(
                p_{s_1}p_{s_2}
                +
                \frac{p_{\alpha}^2- p_{\beta}^2}{s_1s_2}
                \right) \cos(\beta)
                \nonumber\\
        && \qquad 
                -
                \left(
                p_{\alpha}
                \left(\frac{p_{s_2}}{s_1}-\frac{p_{s_1}}{s_2}\right)
                -
                p_{\beta}
                \left(\frac{p_{s_2}}{s_1}
                +
                \frac{p_{s_1}}{s_2}\right)
                \right) \sin(\beta)
                \Bigg]
                +
                (C_1^2-4 p_{\alpha}^2) \cos(\beta) \,.
  \end{eqnarray}
\end{widetext}
Using the inverse mapping for all other canonical variables, these residuals can be written entirely in terms of the second central moments. The moment relations then yield
\begin{eqnarray}
  \label{eq:2}
  R_1 &=& - \sqrt{P} \sin(\alpha + \beta)\,, \\
  R_2 &=& -\sqrt{P} \sin(\alpha - \beta)\,, \\
  R_{12} &=& \sqrt{P} \sin(\alpha)\,,
\end{eqnarray}
where \(P\) is defined in Eq.~(\ref{eq:canonicalSquareRoot}). From these follow
\begin{equation}
  \sin(\alpha) = \frac{R_{12}}{\sqrt{P}}\,,
\end{equation}
\begin{equation}
  \label{eq:10}
  \cos(\alpha)
  =
  \frac{R_2 - R_1}{2 \sqrt{P} \sin(\beta)}\,.
\end{equation}
Eliminating the square root gives
\begin{equation}
  \label{alpha}
  \alpha = \arctan \left(\frac{\sin(\alpha)}{\cos(\alpha)} \right)
  =
  \arctan \left(\frac{2 \sin(\beta) R_{12}}{R_2 - R_1} \right)\,.
\end{equation}
The expression obtained after substituting the second moments is lengthy and not particularly illuminating. An alternative construction of $\alpha$ in terms of second moments is given in Appendix~\ref{sec:altern-deriv-alpha}.

\subsection{Quantum-optical interpretation of the canonical mapping}
\label{sec:reconstr-canon-vari}

In quantum optics, a standard basis for the quadratic
$\mathrm{sp}(4,\mathbb{R})$ generators is given by linear combinations of the
single-mode operators,
\begin{equation}
  \label{quadraticGeneratorsab}
  \hat{a}^2, \enspace
  (\hat{a}^\dagger)^2, \enspace
  \hat{a}^\dagger \hat{a}, \enspace
  \hat{b}^2, \enspace
  (\hat{b}^\dagger)^2, \enspace
  \hat{b}^\dagger \hat{b}, \enspace 
  \hat{a}\hat{b}, \enspace
  \hat{a}^\dagger \hat{b}, \enspace
  \hat{a} \hat{b}^\dagger, \enspace
  \hat{a}^\dagger \hat{b}^\dagger\,,
\end{equation}
or, equivalently, in Hermitian form,
\begin{equation}
  \label{quadraticGeneratorsxp}
  \hat{x}^2, \enspace
  \frac{1}{2} (\hat{x}\hat{p}_x + \hat{p}_x\hat{x}), \enspace
  \hat{p}_x^2, \enspace
  \hat{y}^2, \enspace
  \frac{1}{2} (\hat{y}\hat{p}_y + \hat{p}_y\hat{y}), \enspace
  \hat{p}_y^2, \enspace
  \hat{x}\hat{y}, \enspace
  \hat{x}\hat{p}_y, \enspace
  \hat{p}_x \hat{y}, \enspace
  \hat{p}_x \hat{p}_y\,.
\end{equation}
These operators close under commutation and furnish a realization of $\mathrm{sp}(4,\mathbb{R})$.

Evaluating their expectation values (after centralizing) defines the second moments $\Delta_i = \Delta(\xi^i \xi^j)$ as linear functionals on this Lie algebra. Equivalently, the space of second moments carries the natural Lie-Poisson structure associated with $\mathrm{sp}(4,\mathbb{R})^*$, with
brackets inherited from Eq.~(\ref{Poisson}),
\begin{equation}
  \{\langle \hat{A} \rangle,\langle \hat{B} \rangle\}
  = \langle [\hat{A},\hat{B}] \rangle / i\hbar\,.
\end{equation}
In this sense, the covariance matrix may be identified (after choosing a pairing) with an element of $\mathrm{sp}(4,\mathbb{R})^*$, and the canonical coordinates introduced above provide a Darboux chart on the corresponding symplectic leaves.

Our construction of canonical coordinates may be viewed as a symplectic Gram-Schmidt-type procedure in which the canonical parameters are first adapted to the position sector of the moment space,
\begin{equation}
  \label{eq:28}
  \left(
    \begin{array}{cc}
      \Delta(x^2) & \Delta(xy) \\
      \Delta(xy) & \Delta(y^2)
    \end{array}
  \right)
  =
  \left(
    \begin{array}{cc}
      s_x^2 & s_x s_y \cos(\beta) \\
      s_x s_y \cos(\beta) & s_y^2
    \end{array}
  \right)\,.
\end{equation}
Once this block is fixed, the symplectic structure constrains the mixed and momentum sectors to take correspondingly more nonlinear forms.

This choice explains the systematic appearance of $\sin(\beta)$ in the remaining moment expressions. The parametrization of the position block loses rank at $\beta=0$ and $\beta=\pi$, where $\Delta(xy)=\pm s_x s_y$, and the associated coordinates become singular. Thus, the Darboux chart constructed here is an open chart on the space of second moments.
The degeneracy at $\beta=0$ and $\beta=\pi$ corresponds to maximally correlated or anticorrelated configurations in the position sector. These limits are suggestive of EPR-like correlation regimes at the boundary of the coordinate patch.

As a first step in analyzing the action of the canonical generators, it is useful
to consider the single-mode uncertainty combinations
\begin{equation}
  \label{eq:xyUncertainties}
  U_x = \Delta (x^2)\,\Delta (p_{x}^2) - \Delta (xp_{x})^2\,,
  \qquad
  U_y = \Delta (y^2)\,\Delta (p_{y}^2) - \Delta (yp_{y})^2\,.
\end{equation}
These quantities are no longer invariant under the full two-mode dynamics once
intermode correlations are included. Instead, they become functions on phase space, given by
\begin{align}
  \label{eq:U1}
  U_x &= (p_{\alpha}-p_{\beta})^2 \nonumber\\
      &\quad + \frac{1}{2 \sin^2(\beta)}
       \left[
       (C_1^2-4p_{\alpha}^2)
       - \sqrt{(C_1^2-4p_{\alpha}^2)^2 - (C_1^4-C_2^4)}\,\sin(\alpha+\beta)
       \right],
\end{align}
and
\begin{align}
  \label{eq:U2}
  U_y &= (p_{\alpha}+p_{\beta})^2 \nonumber\\
      &\quad + \frac{1}{2 \sin^2(\beta)}
      \left[
      (C_1^2-4 p_{\alpha}^2)
      - \sqrt{(C_1^2-4p_{\alpha}^2)^2 - (C_1^4-C_2^4)}\,\sin(\alpha-\beta)
      \right].
\end{align}
Although $U_x$ and $U_y$ are no longer constants on the full phase space, they remain independent of the single-mode variables $(s_x, p_{s_x})$ and $(s_y, p_{s_y})$. In particular, they are invariant under the Hamiltonian flows generated by $p_{s_x}$ and $p_{s_y}$ (i.e., single-mode squeezing transformations).

This can be verified explicitly by computing the Hamiltonian flow generated by $p_{s_x}$,
\begin{equation}
  \label{eq:31}
  \frac{{\rm d}}{{\rm d}\lambda}(\cdot) = \{\,\cdot\,, p_{s_x}\}\,,
\end{equation}
under which the position variance $\Delta(x^2)$ increases while the momentum variance $\Delta(p_x^2)$ decreases. Despite this redistribution of fluctuations, the combination $U_x$ remains invariant, reflecting the fact that single-mode squeezing preserves the uncertainty product.
Introducing the logarithmic variable $r_x = \ln s_x$ recovers the standard exponential parametrization of squeezing, making contact with the conventional description of single-mode Gaussian transformations.

The inverse moment formulas for the phase-space generators $p_\alpha$ and $p_\beta$ in Eqs.~(\ref{eq:MomentInversepAlpha}) and (\ref{eq:MomentInverseB}) take the schematic form
\begin{align}
  \label{eq:16}
  p_\alpha
  &\sim
         \frac{1}{2}
         \left[
         \Delta(xy) + \bigl( \Delta(p_x y) - \Delta(x p_y) \bigr)
         \right], \\
  p_\beta
  &\sim
         \frac{1}{2}
         \left[
         \Delta(xy) - \bigl( \Delta(p_x y) + \Delta(x p_y) \bigr)
         \right].
\end{align}
Here we have suppressed the common normalization factor
\[
\sqrt{\Delta(x^2)\Delta(y^2) - \Delta(xy)^2}
= s_x s_y \sin(\beta),
\]
which reflects the choice of coordinates adapted to the position fluctuations, as well as the relative weights of the contributing terms.

The structure of these expressions provides insight into the geometric role of the intermode generators. 
For comparison, the standard Hermitian generator of a two-mode squeezing transformation is \cite[Ch.~5.1]{QuantumContinuous}
\begin{equation}
  \label{two-mode-squeezing}
  \hat{G}_2(r)
  = i r \bigl( \hat{a}^\dagger \hat{b}^\dagger - \hat{a} \hat{b} \bigr)
  = r \bigl( \hat{p}_x \hat{y} + \hat{x} \hat{p}_y \bigr)\,,
\end{equation}
which produces correlated pair excitations and entanglement between the modes. 
At the level of second moments, the symmetric combination $\Delta(p_x y) + \Delta(x p_y)$ associated with this transformation appears explicitly in the definition of $p_\beta$.

The additional terms in $p_\beta$, required for it to be the momentum conjugate to $\beta$ (with $\beta$ fixed by the position correlation $\Delta(xy)$) imply that the transformations generated by $p_\beta$ do not coincide with those generated by $\hat{G}_2$, and that $\beta$ is not identical to the conventional squeezing parameter $r$. Nevertheless, since $p_\beta$ contains the generator of two-mode squeezing, it realizes two-mode squeezing as a Hamiltonian flow on phase space.

In principle, one could instead adapt the canonical chart to the standard two-mode squeezing generator (\ref{two-mode-squeezing}), promoting its parameter $r$ to a canonical coordinate rather than fixing $\beta$ through $\Delta(xy)$. However, enforcing the symplectic structure would then require the remaining generators to take correspondingly more complicated and less transparent forms.

The appearance of the antisymmetric combination $\Delta(p_x y)-\Delta(x p_y)$ in $p_\alpha$ suggests a connection with passive mode-mixing transformations. In the standard optical description, such transformations are generated by the beam-splitter operator \cite[Ch.~5.1]{QuantumContinuous}
\begin{equation}
  \hat G_{\rm BS}(\eta)
  = i\eta\bigl(\hat a^\dagger \hat b-\hat a \hat b^\dagger\bigr)
  = \eta\bigl(\hat p_x \hat y-\hat x \hat p_y\bigr),
\end{equation}
whose defining feature is precisely the antisymmetric cross-quadrature structure
$\hat p_x \hat y-\hat x \hat p_y$.

This correspondence indicates that $p_\alpha$ contains a passive, number-conserving mode-mixing component. However, the identification is not exact. In the present parametrization, the angle $\alpha$ does not enter the position or mixed sectors, but only the momentum block. Consequently, the canonical direction generated by $p_\alpha$ does not coincide with the bare beam-splitter generator in a fixed optical basis.
The expression for $p_\alpha$ also involves the symmetric position correlation $\Delta(xy)$, pointing to an additional correlation-generating contribution. This component can be related to a controlled-phase gate \cite{GaussianStatesOperations}, whose quadratic generator may be written as
\begin{equation}
  \hat G_{\rm CZ}(\phi)
  = \phi\, \hat x \hat y .
\end{equation}

Taken together, these features show that $p_\alpha$ should not be identified with a pure beam-splitter generator. Rather, in the present Darboux chart it corresponds to a dressed canonical direction that combines an antisymmetric, beam-splitter-like component with a symmetric correlation component. The pair $(\alpha,p_\alpha)$ therefore encodes a relative phase-space twist between the modes, which reduces to passive mode mixing in one limit but is not exhausted by the standard passive $\mathrm{u}(2)$ sector alone.

\subsection{Entropy and purity in canonical variables}
\label{sec:entropy-purity}

When restricted to Gaussian states, the two-mode covariance matrix
$\sigma^{ij} = \Delta(\xi^i\xi^j)$ provides a complete parametrization of the quantum state. In this case, the symplectic invariants of $\sigma$ not only characterize
the state but also determine its entropy and purity.

The von Neumann entropy of a quantum state with density matrix $\rho$ is
defined by
\begin{equation}
  \label{eq:5}
  S_V (\rho) = - \mathrm{Tr} \left( \rho \log_2 \rho \right).
\end{equation}
While this definition depends on the full spectrum of $\rho$, for
Gaussian states it admits the closed form
\begin{equation}
  S_V (\sigma) = s_V(2\nu_+/\hbar) + s_V(2\nu_-/\hbar),
\end{equation}
where $\nu_+$ and $\nu_-$ are the symplectic eigenvalues of the
covariance matrix $\sigma$ and
\begin{equation}
  \label{eq:13}
  s_V(\nu) = \frac{\nu+1}{2} \log_2\left( \frac{\nu + 1}{2}\right)
  - \frac{\nu - 1 }{2} \log_2 \left( \frac{\nu - 1}{2}\right).
\end{equation}

An alternative measure of mixedness is the purity $\mu(\rho)$,
\begin{equation}
  \mu(\rho) = \mathrm{Tr} (\rho^2),
\end{equation}
which for Gaussian states takes the form
\begin{equation}
  \label{eq:8}
  \mu(\sigma) = \frac{\hbar^2/4}{\sqrt{\det(\sigma)}}.
\end{equation}

Using Eqs.~(\ref{eq:quadraticCasimirC12}) and (\ref{eq:quarticCasimirC22}), one finds
\begin{equation}
  \label{eq:nuCasimirRelation}
  \nu_\pm^2 = \frac{1}{2}\left(C_1^2 \pm C_2^2\right).
\end{equation}
Consequently, both the entropy and the purity can be expressed entirely in terms of the Casimirs.
In particular, since $\det(\sigma)=\nu_+^2 \nu_-^2$, one obtains
\begin{equation}
  \label{eq:detSigmaCasimir}
  \det(\sigma) = \frac{1}{4}\left(C_1^4 - C_2^4\right),
\end{equation}
so that the purity becomes
\begin{equation}
  \label{eq:purityCasimir}
  \mu(\sigma)
  =
  \frac{\hbar^2}{2\sqrt{C_1^4 - C_2^4}}.
\end{equation}
The bounds on $C_1$ and $C_2$ discussed in
Sec.~\ref{sec:inverse-mapping} ensure that the purity satisfies
$\mu(\sigma)\leq 1$, as required for a physical quantum state.

Through these relations, $C_1$ and $C_2$, invariant under all Hamiltonian evolutions, provide a global information-theoretic characterization of the quantum state, encoding its mixedness and entropy.

While one may also construct local or nonconserved information measures on phase space, these are not invariant under the full $\mathrm{sp}(4,\mathbb{R})$
dynamics. In particular, entanglement, as quantified by the logarithmic negativity constructed from the smallest partially transposed symplectic eigenvalue $\nu_-^{\rm PT}$, can vary dynamically. It is expressed as a function on phase space and  in the companion paper~\cite{EntropyPurity}.
In what follows, we focus on the global invariants and their role in shaping the geometric character of the theory.

\subsection{Riemannian limits and Gaussian states}
\label{sec:riem-spec-cases}

The Finsler structure defined by the quantum-corrected geodesic Hamiltonian, Eq.~(\ref{eq:quarticFinslerHamiltoniana}), will define a Riemannian geometry if the quartic part of it, Eq.~(\ref{eq:quarticPart}), is reducible.

\subsubsection{Gaussian states}
\label{sec:case-1}

The statistical moments \(\Delta(x^\alpha p^\beta)\) (for all multi-indices \(\alpha\) and \(\beta\)) provide a complete
parametrization of \(N\)-mode quantum states, pure or mixed. However, pure
Gaussian states are uniquely characterized by a restricted subset of the
second-order moments. In the position representation, Gaussian states are
fully specified by their position-position statistics \(\Sigma_x^{ij} = \Delta(x^ix^j)\) and
position-momentum statistics \(\Sigma_{xp}^{ij} = \Delta(x^ip^j)\). The second-order momentum
statistics, as well as all higher-order moments, are entirely determined by
these two sets of parameters.

The \(N\)-mode pure Gaussian is parametrized explicitly by the position-space
wave function
\begin{equation}
  \label{eq:twomodeGaussianWaveFunction}
  \psi_G(x,\Sigma_x,\Sigma_{xp})
  = \left( \frac{1}{2\pi \sqrt{\det{\Sigma_x}}}\right)^{1/2}
  \exp\left(
  - \frac{1}{4}x^T \Sigma_x^{-1}\left(\mathbb{I} - i \frac{\Sigma_{xp}}{\hbar/2}\right)x
  \right)
\end{equation}
where \(x = (x_1,\ldots, x_N)\) is the position vector and matrix multiplication is assumed. The modulus of Eq.~(\ref{eq:twomodeGaussianDensity}) recovers the standard Gaussian density
\begin{equation}
  \label{eq:twomodeGaussianDensity}
  p_G(x,\Sigma_x)
  = \frac{1}{2\pi \sqrt{\det{\Sigma_x}}}
  \exp\left(
  - \frac{1}{2}x^T \Sigma_x^{-1}x
  \right).
\end{equation}

The momentum-space wave function is obtained as the Fourier transform of the
position-space Gaussian wave function. This transformation can be performed analytically using the
standard Gaussian integral formula
\begin{equation}
  \label{eq:36}
  \int \exp\left(-\frac{1}{2} x^T A  x + J^T  x\right) \mathrm{d}^nx
  =
  \sqrt{\frac{(2\pi)^n}{\det{A}}}
  \exp\left(\frac{1}{2} J^T A^{-1} J \right)\,.
\end{equation}
Applying this formula, the resulting Gaussian wave function in momentum space is
\begin{eqnarray}
  \label{eq:37}
  &&\tilde{\psi}_G(p,\Sigma_x,\Sigma_{xp})
  =
  \left(
    \frac{\det{(\Sigma_x)}}
    {(2\pi)^2\det{\left(\frac{\hbar^2}{4}\mathbb{I} + \Sigma_{xp}^2\right)}}
      \right)^{1/4}\\
  &&\times\exp\left(
    - \frac{1}{4} p^T
    \left(\frac{\hbar^2}{4}\mathbb{I} + \Sigma_{xp}^2\right)^{-1}
    \left(\mathbb{I} + i \frac{\Sigma_{xp}}{\hbar/2}\right)
    \Sigma_x
    p
    + \frac{i}{2}
    \arctan{\left(
        \frac{\frac{\hbar}{2} \mathrm{Tr} (\Sigma_{xp})}
        {\frac{\hbar^2}{4} - \det{(\Sigma_{xp})}}
      \right)}
    \right)\,. \nonumber
\end{eqnarray}

From this expression, it follows that for Gaussian states, the momentum covariance matrix is fully determined by the matrix equation
\begin{equation}
  \label{eq:SigmapEq}
  \Sigma_p = \Sigma_x^{-1} \left(\frac{\hbar^2}{4}\mathbb{I} + \Sigma_{xp}^2\right)
\end{equation}
or equivalently, by the equations
\begin{equation}
  \label{eq:matrixUncertaintyRel}
  \Sigma_x\Sigma_p - \Sigma_{xp}^2 = \frac{\hbar^2}{4}\mathbb{I}.
\end{equation}
This result generalizes to multiple modes the statement that Gaussian states saturate the quantum uncertainty relation.

Specializing to the two-mode case, either Eq.~(\ref{eq:SigmapEq}) or Eq.~(\ref{eq:matrixUncertaintyRel}) represents a system of four nonlinear constraint equations. When expressed in canonical coordinates, these equations impose the following constraints on the canonical variables
\begin{gather}
  C_1^2 = \frac{\hbar^2}{2},   \label{eq:Gauss1} \\
  C_2 = 0,   \label{eq:Gauss2} \\
  p_\alpha = 0.   \label{eq:Gauss3}
\end{gather}
The detailed derivation of these results is presented in Appendix~\ref{a:gaussian}. When translated back to the original moment parametrization, these three conditions indicate that only seven of the ten second-order moments for the two-modes are independent. Typically, the independent set is taken to include the three position-related moments \(\Sigma_x\) and the four mixed position-momentum moments \(\Sigma_{xp}\).

When Eqs.~(\ref{eq:Gauss1})--(\ref{eq:Gauss3}) are satisfied (or so long as $C_2$ and $p_{\alpha}$ are much smaller than $C_1$ and $p_{\beta}$), then
the quartic part in the Hamiltonian (\ref{eq:quarticFinslerHamiltoniana}) is
negligible. In this case, we can approximate from Eqs.~(\ref{eq:U1}) and
(\ref{eq:U2}),
\begin{equation}
  U_x\approx U_y\approx p_{\beta}^2+\frac{C_1^2}{2\sin^2\beta}
\end{equation}
and the quantum contributions to $p_x^2$ and $p_y^2$ from momentum variances
are
\begin{equation}
  \label{eq:Deltapx2Gauss}
  \Delta(p_x^2)=p_{s_x}^2+ \frac{p_{\beta}^2}{s_x^2}
  +\frac{C_1^2}{2 s_x^2\sin^2{(\beta)}},
\end{equation}
\begin{equation}
  \label{eq:Deltapy2Gauss}
  \Delta(p_y^2)=p_{s_y}^2
  + \frac{p_{\beta} }{s_y^2} 
  + \frac{C_1^2}{2s_y^2\sin^2(\beta)},
\end{equation}
and
\begin{eqnarray}
  \label{eq:Deltap1p2Gauss}
  \Delta(p_x p_y)
  &=& \left(p_{s_x}p_{s_y} - \frac{p_{\beta}^2}{s_xs_y} -\frac{C_1^2}{2\sin^2(\beta) s_x s_y}\right)\cos(\beta) \\
  && - \sin(\beta)p_{\beta}\left(\frac{p_{s_y}}{s_x}+ \frac{p_{s_x}}{s_y}\right)\,. \nonumber
\end{eqnarray}

These expressions indicate that two-mode pure Gaussian states are fully characterized on the six-dimensional phase space coordinatized by \((s_x,p_{s_x};s_y,p_{s_y};\beta, p_\beta)\). On this space, the moment expressions (\ref{eq:Deltapx2Gauss})--(\ref{eq:Deltap1p2Gauss}) can be interpreted as centrifugal kinetic energies in two auxiliary three-dimensional configuration spaces (one for each mode). This subspace arises from reducing the full ten-dimensional space of second moments by the purity constraints Eqs.~(\ref{eq:SigmapEq}), which fix the values of the Casimirs \(C_1\) and \(C_2\), and eliminate the degree of freedom associated with the conjugate pair \((\alpha,p_\alpha)\).

\subsubsection{Case 2}

In the pure Gaussian case, where \(\nu_+ = \nu_- = \hbar/2\) and \(p_\alpha=0\), the Finsler root \(\sqrt{P}\) vanishes identically, and the resulting two-mode geometry reduces to a Riemannian structure. This represents a special limiting configuration where the state is both pure and symplectically isotropic.

Riemannian geometry can also emerge in more general, nonpure settings. Specifically, even when the symplectic eigenvalues take nonlimiting values, the geometry remains Riemannian if \(p_\alpha\) can be chosen so that the Finsler discriminant vanishes. However, the momentum \(p_\alpha\) is not free to vary arbitrarily. For given values of the Casimirs \(C_1\) and \(C_2\), reality conditions on the Finsler discriminant imposes a nontrivial constraint
\begin{equation} \label{eq:canonicalSquareRootIneq}
  P(p_\alpha,C_1,C_2) = (C_1^2 - 4p_{\alpha}^2)^2- (C_1^4-C_2^4) \ge 0.
\end{equation}
This condition carves out a forbidden interval around \(\vert p_\alpha\vert = C_1/2\). Specifically, if \( \vert p_\alpha \vert < C_1 / 2\), then Eq.~(\ref{eq:canonicalSquareRootIneq}) requires
\begin{equation}
  \label{eq:palphaBound1}
  0
  \le
  \vert p_\alpha \vert
  \le \sqrt{\frac{C_1^2-\sqrt{C_1^4-C_2^4}}{4}}
  = \frac{\nu_+ - \nu_-}{2} \,.
\end{equation}
On the other hand, if \(\vert p_\alpha\vert > C_1/2\), the inequality becomes
\begin{equation}
  \label{eq:palphaBound2}
  \vert p_\alpha \vert \ge \sqrt{\frac{C_1^2+\sqrt{C_1^4-C_2^4}}{4}}
  = \frac{\nu_+ + \nu_-}{2} \,.
\end{equation}

Reality conditions thus imply that the allowed values of \(p_\alpha\) lie outside the exclusion band defined by
\begin{equation}
  \label{eq:palphaBoundnu}
  p_- := \frac{\nu_+ - \nu_-}{2}
  < \vert p_\alpha \vert < \frac{\nu_+ + \nu_-}{2}
  =: p_+\,.
\end{equation}
% bounded by the minimal and maximal angular momenta \((p_-, p_+)\) with
% \begin{gather} \label{eq:palphaBoundnu}
%   p_- = \frac{\nu_+ - \nu_-}{2}\,, \\
%   p_+ = \frac{\nu_+ + \nu_-}{2}\,.
% \end{gather}
In the extremal cases where \(p_\alpha\) assumes a boundary value, \(p_\alpha \ne 0\), yet the Finsler root still vanishes, signaling a return to a Riemannian form despite the presence of nontrivial correlations. These configurations generalize the pure-state limit and reveal a broader class of geometrically special states.

%saturates either the upper or lower bounds allowed by the reality condition:
% \begin{equation}\label{eq:absIneq}
%   \vert C_1^2 - 4 p_\alpha^2 \vert \ge \sqrt{C_1^4 - C_2^4}.
% \end{equation}

% \begin{equation}
%   \frac{\nu_+ - \nu_-}{2} \le \vert p_\alpha \vert \le \frac{\nu_+ + \nu_-}{2}
% \end{equation}

Explicitly, when \(p_\alpha\) saturates the lower bound \(p_\alpha = p_-\), we have the momentum variances
\begin{equation}
  \label{eq:Deltapx2Gauss2}
  \Delta(p_x^2)=p_{s_x}^2+ \frac{(p_--p_{\beta})^2}{s_x^2}
  +\frac{\nu_+\nu_-}{s_x^2 \sin^2{(\beta)}}\,,
\end{equation}
\begin{equation}
  \label{eq:Deltapy2Gauss2}
  \Delta(p_y^2)=p_{s_y}^2
  + \frac{(p_-+p_{\beta})^2 }{s_y^2} 
  + \frac{ \nu_+\nu_-}{s_y^2\sin^2(\beta)}\,,
\end{equation}
and
\begin{eqnarray}
  \label{eq:Deltap1p2Gauss2}
  \Delta(p_x p_y)&=&\left(p_{s_x}p_{s_y} 
                     +\frac{(p_-)^2-p_{\beta}^2}{s_xs_y}
                     -\frac{\nu_+\nu_-}{s_x s_y\sin^2(\beta)}
                     \right)\cos(\beta) \\
                && +\left(
                     \frac{p_{s_y}}{s_x}(p_--p_{\beta})
                     - \frac{p_{s_x}}{s_y}(p_{\beta}+p_-)\right)\sin(\beta)\,.
\end{eqnarray}
When \(p_\alpha\) saturates the upper bound \(p_\alpha = p_+\) the azimuthal terms become negative but otherwise the terms retain the same structure.

The presence of a forbidden interval, \((p_-, p_+)\), through which \(p_\alpha\) cannot evolve under any Hamiltonian flow might be an artifact of our truncation. The variables \(p_\alpha, C_1,\) and \(C_2\) are of order \(\hbar\) but the discriminant \(P(p_\alpha,C_1,C_2)\) is quartic. The discriminant contributes at quadratic order because of the square root.
It is plausible that in a higher-order truncation, fourth-order variables could modify the discriminant and the reality condition (\ref{eq:canonicalSquareRootIneq}) to connect the allowed regions.
If the forbidden interval is not an artifact, then it is intriguing to consider its implications for the second-order theory.

According to reality conditions, the quantum state structure directly determines not just the qualitative nature of the geometry, but also the accessible phase space. The bounds on \(p_\alpha\) show that its range is dictated by the symplectic eigenvalues \(\nu_+\) and \(\nu_-\), which encode the quantum uncertainty of the system. If the state is prepared with \(\vert p_\alpha \vert \ge p_+\), then its sign remains fixed under unitary evolution. In contrast, within the inner band, sign changes are allowed. This asymmetry introduces a form of dynamical irreversibility, or even topological protection, wherein the phase-space structure imposes global constraints on state evolution. Transitions between disjoint \(p_\alpha\) sectors are dynamically forbidden---not by energetic barriers, but by deeper algebraic and geometric constraints.

\subsubsection{Case 3}
\label{sec:case-3}

Another quadratic case is obtained if $C_1=C_2$ with arbitrary $p_{\alpha}$,
such that
\begin{equation}
  U_x=(p_{\alpha}-p_{\beta})^2-\frac{1}{2}|C_1^2-4p_{\alpha}^2|
  \frac{\sin(\alpha+\beta)}{\sin\beta}
\end{equation}
and a similar expression for $U_y$.
The
positivity condition on (\ref{eq:detsigma}) implies $|C_1|>|C_2|$. This Riemannian case is therefore forbidden for physical states. This demonstrates how information-theoretic quantities inherently shape the Finslerian geometry experienced by the system.

% A Gaussian state is the ground or thermal state of a quadratic Hamiltonian. The Hamiltonian controlling the statistics' evolution (\ref{eq:Hq2dof}) is, at this approximation not a quadratic function, but homogeneous of degree two.

% Q1: is the ground state of (\ref{eq:Hq2dof}) Gaussian?

% A definition of purity for beyond-Gaussian states from the covariance matrix. We construct the harmonic oscillator state.

\section{Time dilation}
\label{sec:time-dilation}

We now return to the notion of quantum proper time. For particle motion in two spatial dimensions---or equivalently for a clock realized by two bosonic modes---the Hamiltonian expression (\ref{eq:HamiltonianProperTime}) provides a complete characterization of proper time. 

In many geometric formulations, however, proper time is expressed as a functional on the tangent bundle. To make contact with this perspective, we now derive the corresponding Lagrangian formulation.

The convexity of the quartic Hamiltonian
(\ref{eq:quarticFinslerHamiltoniana}) guarantees the existence of its
Legendre transform. However, the resulting velocity-momentum relation is cubic and cannot be inverted in closed form. 
Nonetheless, the quantum moment framework is naturally perturbative. A perturbative Legendre transform, carried out to the same order of accuracy as the Hamiltonian, yields the following expression for the Lagrangian:
\begin{equation}
  L(x,\dot{x})
  =
  -\frac{m}{2}\,
  B_{\bar{a}\bar{b}}\dot{x}^{\bar{a}}\dot{x}^{\bar{b}}
  + \frac{m c^2}{2} 
  \pm \frac{m}{2}
  \sqrt{
    A_{\bar{a}\bar{b}\bar{c}\bar{d}}
    \dot{x}^{\bar{a}}\dot{x}^{\bar{b}}\dot{x}^{\bar{c}}\dot{x}^{\bar{d}}
    +
    \frac{
      \left(
        \dot{\varphi}_2
        \sqrt{
          A_{\bar{a}\bar{b}\bar{c}\bar{d}}
          \dot{x}^{\bar{a}}\dot{x}^{\bar{b}}\dot{x}^{\bar{c}}\dot{x}^{\bar{d}}
        }
      \right)^{\!4/3}
    }{
      (A^{C_2C_2C_2C_2})^{1/3}
    }
  }\,.
  \label{eq:Lagrangian-full}
\end{equation}

Here $B_{\bar{a}\bar{b}}$ denotes a pseudo-inverse of the rank-deficient
quadratic form $B^{\bar{a}\bar{b}}$. The tensor
$A_{\bar{a}\bar{b}\bar{c}\bar{d}}$ is obtained by lowering indices on
$A^{\bar{a}\bar{b}\bar{c}\bar{d}}$ using $B_{\bar{a}\bar{b}}$. The variable
$\varphi_2$ is conjugate to the Casimir $C_2$, satisfying
$\{\varphi_2,C_2\}=1$. The overall sign of the quartic contribution is fixed by the corresponding sign in the Hamiltonian, determined by the contraction in Eq.~(\ref{purityWeightSign}).

The second term under the square root arises at this order to account for
the contribution of the second Casimir $C_2$ to the time dilation, which is
not captured by the quadratic form $B_{\bar{a}\bar{b}}$. For simplicity in
the following discussion, we neglect this contribution by setting
$\dot{\varphi}_2 = 0$, thereby isolating the effect of the remaining quantum
degrees of freedom.

Imposing the Lagrangian constraint $L = m c^2$, dividing by $\dot{t}^2$, and
rearranging yields the quantum-corrected time-dilation formula
\begin{equation}
  \label{eq:6}
  \frac{{\rm d} t}{{\rm d} \tau}
  =
  \frac{1}{\sqrt{\left(
    - B_{\bar{a}\bar{b}}
    \frac{\dot{x}^{\bar{a}} \dot{x}^{\bar{b}}}{\dot{t}^2}
    \pm \frac{\sqrt{A}}{\dot{t}^2}
  \right) / c^2}} \,.
\end{equation}
In terms of deparametrized velocities
$v^{\bar{a}} := \dot{x}^{\bar{a}}/\dot{t}$, this becomes
\begin{equation}
  \label{eq:dtdtau-QEffective}
  \frac{{\rm d} t}{{\rm d} \tau}
  =
  \frac{1}{\sqrt{\left(
    - B_{\bar{a}\bar{b}} v^{\bar{a}} v^{\bar{b}}
    \pm \sqrt{
      A_{\bar{a}\bar{b}\bar{c}\bar{d}}
      v^{\bar{a}} v^{\bar{b}} v^{\bar{c}} v^{\bar{d}}
    }
  \right) / c^2}}\,.
\end{equation}
This expression makes explicit how non-Riemannian (quartic) corrections modify the familiar time-dilation factor.

The expression can be written more compactly by defining a Finsler norm
\begin{equation}
  \label{eq:21}
  F(v) =
  \sqrt{
    - B_{\bar{a}\bar{b}} v^{\bar{a}} v^{\bar{b}} \pm \sqrt{A_{\bar{a}\bar{b}\bar{c}\bar{d}} v^{\bar{a}}v^{\bar{b}}v^{\bar{c}}v^{\bar{d}}}
  }\,.
\end{equation}
In terms of this norm, the proper time along a trajectory with tangent \(v\) is
\begin{equation}
  \label{eq:34}
  \tau = \frac{1}{c} \int_0^1 F(v) \, {\rm d}t \,,
\end{equation}
a direct generalization of the Riemannian distance functional.

The quartic correction is genuinely non-Riemannian. It arises only when the clock is in a quantum state with spread over multiple phase-space directions. Physically, it encodes how fluctuations of the clock’s internal degrees of freedom modify its effective tick rate.

In the near-Riemannian regime, the quartic term is small, so the time-dilation factor reduces smoothly to the classical result. But as the quantum correlations increase, the correction perturbs the Lorentz factor, effectively broadening the dependence of proper time on velocity and position. In other words, the clock no longer measures time as if it were a sharp worldline, but rather as if it were an extended object in phase space.

The special case of a stationary clock, defined by
$p_x = p_y = p_{s_x} = p_{s_y} = p_\alpha = p_\beta = 0$,
isolates the contribution of the irreducible quantum fluctuations
$\nu_\pm$ to the effective clock rate and is considered in the companion paper~\cite{EntropyPurity}.

\section{Discussion}
\label{sec:discussion}

In curved spacetime, a particle's motion is typically described by geodesics
of a quadratic metric. Corresponding observables---for instance based on particle
dispersion relations---are tightly constrained by experimental data. However,
quantum effects introduce deviations from these classical trajectories. Using
the Hamiltonian formulation of the geodesic problem, applying canonical quantization,
and parametrizing the state through moments of the canonical operators,
this paper has shown that the quantum-corrected dynamics of a freely falling
particle retain a Hamiltonian structure. To fully utilize this
structure, a faithful mapping of the second-order moments to canonical
variables enabled us to distinguish quantum variables as either configuration
variables or momentum variables and to establish the kinetic
component of the Hamiltonian in a systematic way.

The identification of these canonical variables is further enriched by quantum
information-theoretic considerations. Standard results for Gaussian states
explicitly link entropy and purity to the covariance matrix of second-order
statistics. Our methods, which are built on the semiclassical approximation
(\ref{eq:hierarchy}) and which apply also for non-Gaussian or mixed states,
encode these measures of quantum coherence and correlations into the Casimir's
\(C_1\) and \(C_2\) of the canonical mapping previously derived in
\cite{Bosonize,EffPotRealize}. Through this chain of mappings and their
inversion analyzed here, these quantum information-theoretic quantities not
only characterize the state but also influence the particle's interaction with
the underlying geometry.

The four dimensions of the spacetime metric \(g_{ab}\) originate from the four-dimensional tangent space of the classical point particle. However, it has been argued that physical observables in canonical quantum gravity can only be defined by considering the quantum properties of the material bodies that form the reference systems \cite{QFR,GeomObs2,ReferenceFrames}. 
As shown in this work, when two classical degrees of freedom of a particle used as a reference frame are quantized, the classical tangent space is effectively extended to a ten-dimensional space. Quantum reference frames therefore necessarily require a deeper understanding of spacetime structure, extending beyond the conventional four-dimensional tangent space.

In this extended framework, the moments \(\Delta(p_x^2)\), \(\Delta(p_y^2)\), and \(\Delta(p_xp_y)\) are no longer quadratic functions of the tangent space, making it impossible to describe the quantum dynamics within the standard spacetime setting provided by Riemannian geometry, even extended to additional dimensions. Nonetheless, these moments remain homogeneous functions of degree two in new momenta, making Finsler geometry a suitable broader framework that incorporates quantum degrees of freedom while preserving reparametrization invariance.

The full Finsler Hamiltonian for the two-mode system, given by Eq.~(\ref{eq:quarticFinslerHamiltoniana}), simultaneously encodes the spacetime curvature experienced by the first and second moments, as well as their mutual backreaction. Because the complete, untruncated quantum Hamiltonian (\ref{eq:Hq}) preserves the same homogeneity properties as the classical Hamiltonian at any order, we anticipate that any physical reference system constructed using quantum continuous variables will experience spacetime through a Finsler geometry analogous to the one described here.

The quantum-corrected dispersion relation, given by Eq.~(\ref{eq:quantumDispersionWeak}), expressed in terms of an effective
mass, provides a simpler framework for bridging quantum mechanics and general
relativity. By associating quantum corrections with an effective mass, the
particle's trajectory is modified as if influenced by additional mass or
energy linked to its quantum state. This effective mass directly connects
quantum state properties, such as spread, to the particle's interaction with
spacetime.

Our results also extend the framework of quantum information theory beyond its connection to gravity or spacetime geometry. While the standard quadratic optical generators (\ref{quadraticGeneratorsab}) correspond to experimentally accessible transformations, their underlying canonical symplectic structure becomes manifest only after a mapping to canonical variables such as the one used here. The present formalism thus complements the experimental program by revealing the nontrivial geometric structure that governs quantum state dynamics.

The theory's predictions could be tested in several experimental settings,
particularly in regimes where quantum coherence and spacetime curvature are
both significant. Quantum coherence effects on massive particles, such as in
cavity optomechanics using either silica nanospheres or cold atomic ensembles
\cite{Optomechanics}, could probe the effective mass corrections predicted by
the theory. By subjecting ultracold atoms in coherent superpositions to
gravitational potentials or artificial spacetime geometries created using
optical lattices, deviations from classical trajectories predicted by the
theory could be observed. If states can be controlled in such experiments, the
influence of all the quantum parameters discussed here can be analyzed.

\section*{ACKNOWLEDGEMENTS}

\noindent The authors thank Domenico Giulini and Christian Pfeifer for
interesting discussions. This work was supported in part by NSF Grant No. PHY-2206591.

\appendix

\section{Block entries of the two-mode quadratic tensor}
\label{a:block-entries-two}

When needed, we denote the basis of the tangent space \(V\cong \mathbb{R}^{10}\) spanned by the extended momentum covector by \(e_{\bar{a}}\) and the dual basis of \(V^*\) by \(\vartheta^{\bar{a}}\) where the \(\bar{a}\) runs from 1 to 10 in both cases. As usual, these satisfy the duality condition
\begin{equation}
  \label{eq:22}
  e_{\bar{a}}(\vartheta^{\bar{b}}) = \vartheta^{\bar{b}}(e_{\bar{a}}) = \delta^{\bar{b}}_{\bar{a}}\,.
\end{equation}

The componentwise expectation value of the metric is in the two-mode case a two-variable Taylor series with
\begin{eqnarray}  \label{eq:9b}
  &&\langle g^{ab}(\hat{x}, \hat{y}) \rangle\nonumber\\
  &=& \left(
      1
      + \frac{1}{2} \left(
        s_x^2 \partial_x^2 +2 s_x s_y \cos (\beta ) \partial_{xy}^2  + s_y^2 \partial_y^2 \right)
    \right)
      g^{ab}(x) + O(\hbar^{3/2}) \\
  &=& \left(
      1
      + \frac{1}{2} \left(
        \Delta(x^2) \partial_x^2 +2 \Delta(xy) \partial_{xy}^2  + \Delta(y^2) \partial_y^2 \right)
    \right)
    g^{ab}(x) + O(\hbar^{3/2})
\end{eqnarray}
which represents the same state-averaging of spacetime as in the 1-degree-of-freedom case.

The \(2\times 4\) block \(D[g^{ib}]\) defined for \(i = 1,2\) and \(b = t,x,y,z\) is a material derivative specifying the inverse metric coefficients for the off-diagonal terms in the \(p_{s_i}\) and \(p_b\) dimensions with
\begin{equation}
  \label{eq:19}
  D[g^{ib}]
  =
  \begin{pmatrix}
    (s_x \partial_x +s_y \cos (\beta ) \partial_y) g^{xb} \\
    (s_y \partial_y +s_x \cos (\beta ) \partial_x) g^{yb}
  \end{pmatrix}.
\end{equation}
This block is a generalization of the corresponding expression for 1 degree
of freedom. In equations of motion, when the inverse-metric components
\(g^{xb}\) or \(g^{yb}\) vary along a classical coordinate direction \(x\) or
\(y\), this expression will couple fluctuation momenta \(p_{s_x}\) or
\(p_{s_y}\) to \(p_x\) or \(p_y\), respectively. In other words, nonzero
coefficients in \(D[g^{ib}]\) indicate that the cotangent vectors associated
with classical momenta and fluctuation momenta are not orthogonal.

The \(2\times 2\) block \(\tilde{g}^{ij}\) defined for \(i,j=1,2\) specifies
the inverse-metric coefficients for \(p_{s_i}\) and \(p_{s_j}\)
as
\begin{equation}
  \label{eq:20}
  \tilde{g}^{ij}
  =
  \begin{pmatrix}
    g^{xx} & \cos (\beta) g^{xy} \\
    \cos (\beta) g^{xy} & g^{yy}
  \end{pmatrix}\,.
\end{equation}
The determinant of \(\tilde{g}^{ij}\) is
\begin{equation}
  \label{eq:17}
  \det(\tilde{g}^{ij}) = g^{xx}g^{yy} - \cos^2(\beta) (g^{xy})^2,
\end{equation}
indicating that the shape of the metric ellipsoid for the quantum correlations \(s_x\), \(s_y\) differs from that of the classical coordinates \(x\) and \(y\) when \(\Delta(xy)\ne 0\).

In the \((xp)\) ordering defined by
\begin{equation}
  \label{eq:29}
  x_i = (x,y,p_x,p_y),
\end{equation}
the covariance matrix is rewritten as
\begin{equation}
  \label{eq:4}
  \Sigma = 
  \left(
    \begin{array}{cc}
      \Sigma_x & \Sigma_{xp} \\
      \Sigma_{xp}^\top & \Sigma_p
    \end{array}
  \right)
\end{equation}
so that we have position-position, position-momentum, and momentum-momentum covariance matrices
\begin{equation}
  \label{eq:Sigma}
  \Sigma_x =
  \begin{pmatrix}
    \Delta(x^2) & \Delta(xy) \\
    \Delta(xy) & \Delta(y^2)
  \end{pmatrix}, \quad
  \Sigma_{xp} =
  \begin{pmatrix}
    \Delta(xp_x) & \Delta(xp_y) \\
    \Delta(p_x y) & \Delta(yp_y)
  \end{pmatrix}, \quad
  \Sigma_{p} =
  \begin{pmatrix}
    \Delta(p_x^2) & \Delta(p_xp_y) \\
    \Delta(p_x p_y) & \Delta(p_y^2)
  \end{pmatrix}.
\end{equation}
The quadratic Casimir of the two-mode mapping, \(C_1^2\), generalizes the quadratic Casimir of the single-mode mapping, \(U\). Here, the inverse metric coefficient in the \(e_{C_1} \otimes e_{C_1}\) subspace equals half the trace of the inverse metric against the inverse position-covariance matrix
\begin{equation}
  \mathrm{Tr}\left(\Sigma_x^{-1} g^{-1}\right) = \csc ^2(\beta )
  \left(\frac{g^{xx}}{s_x^2}+\frac{g^{yy}}{s_y^2} - \frac{2 \cos (\beta ) g^{xy}}{s_xs_y}\right).
\end{equation}

Analogs of the cross-mode momenta \(p_\alpha\) and \(p_\beta\) did not appear in the one-dimensional
restriction. Their role in the effective geometry described by
(\ref{eq:gQinv2dof}) arises due to the blocks \(E\), \(C\), and \(\Theta\).

The term \((\partial g^{aj}/\partial x^k) p_a\Delta(x^kp_j)\) couples \(p_\alpha\) and \(p_\beta\) to the classical momenta \(p_a\) through antisymmetric and symmetric derivatives of the metric
\begin{equation}
  \label{eq:18}
  E^T = \sin (\beta ) 
  \begin{pmatrix}
    (s_y/s_x) \partial_y g^{xb} - (s_x/s_y) \partial_x g^{yb} \\
    -\left((s_y/s_x) \partial_y g^{xb} + (s_x/s_y) \partial_x g^{yb}\right)
  \end{pmatrix}\,.
\end{equation}
The components of \(e_\beta \otimes e_b\) (the lower row) are symmetric gradients that capture how a quantum
state interprets stretching or compression of the spacetime, without
introducing rotation or shearing. That is, quantum correlations may affect how
nearby geodesics converge or diverge. If the spacetime metric has vorticity, then components of
\(e_\alpha \otimes e_b\) are nonzero and couple the classical degrees of
freedom directly to \(p_\alpha\). Such conditions are related to the
development of non-Gaussianity through $p_{\alpha}$, as shown in Sec.~\ref{sec:riem-spec-cases}.

The coefficients of \(e_{s_i}\otimes e_\alpha\) and
\(e_{s_i}\otimes e_\beta\) arise entirely from the term \(g^{xy}\Delta(p_xp_y)\) and contribute the block
\begin{equation}
  \label{eq:3}
  C = \sin(\beta)\frac{g^{xy}}{s_xs_y}
  \begin{pmatrix}
    -s_x
    & -s_x\\
    s_y
    & -s_y
  \end{pmatrix}\,.
\end{equation}

Finally, the block containing the coefficients of the angular coordinates is
\begin{equation}
  \label{eq:3b}
  \Theta =
  \begin{pmatrix}
    \sin^2(\beta) \mathrm{Tr}\left(\sigma_3\Sigma_x^{-1}\sigma_3 g^{-1}\right)
      - 2 \mathrm{Tr}\left(\Sigma_x^{-1} g^{-1}\right)
    & -g^{xx}/s_x^2 + g^{yy}/s_y^2 \\
    -g^{xx}/s_x^2 + g^{yy}/s_y^2
    & \sin^2(\beta) \mathrm{Tr}\left(\Sigma_x^{-1} g^{-1}\right)
  \end{pmatrix}
\end{equation}
where 
\begin{eqnarray}
  \label{Pi}
  \sigma_3
  = \left(
  \begin{array}{cc}
    1 & 0 \\
    0 & -1
  \end{array}
  \right)\,.
\end{eqnarray}

\section{Alternative derivation of $\alpha$}
\label{sec:altern-deriv-alpha}

An alternative expression for the dependence of $\alpha$ on second moments from that obtained in the main text as Eq.~(\ref{alpha}) can be obtained by computing $U_x-U_y$ in two
different ways, first from (\ref{eq:xyUncertainties}) and then from (\ref{eq:U1}) and
(\ref{eq:U2}), such that
\begin{equation} \label{alphaDelta}
  \Delta(x^2)\Delta(p_{x}^2)- \Delta(y^2)\Delta(p_{y}^2)-
  \Delta(xp_{x})^2+\Delta(yp_{y})^2= -4p_{\alpha}p_{\beta}- \sqrt{P}
  \frac{\cos\alpha}{\sin\beta}\,.
\end{equation}
The remaining equations in
(\ref{eq:MomentInverseA})--(\ref{eq:MomentInverseB}) as well as
(\ref{eq:quadraticCasimir}) and  (\ref{eq:12}) then provide a complete
expression for $\cos\alpha$ in terms of second-order moments. In particular,
the terms on the left-hand side of (\ref{alphaDelta}) as well as the product
$p_{\alpha}p_{\beta}$ on the right can all be written in terms of different
uncertainty products, for which we use the general definition
\begin{equation}
  \Delta_{A_1A_2}^{B_1B_2}=\Delta(A_1B_1)\Delta(A_2B_2)-\Delta(A_1B_2)\Delta(A_2B_1)\,.
\end{equation}
With this notation, we can write
\begin{equation} \label{a:alpha}
  \cos\alpha=-\frac{\sin\beta}{\sqrt{P}}\frac{(\Delta_{yp_y}^{xy})^2-
    (\Delta_{xp_y}^{xy})^2+\Delta_{xy}^{xy}\Delta_{xp_x}^{xp_x}-
    \Delta_{xy}^{xy}\Delta_{yp_y}^{yp_y}}{\Delta_{xy}^{xy}} \,.
\end{equation}

\section{Solution of the pure Gaussian constraints}
\label{a:gaussian}

The diagonal equations of Eqs.~(\ref{eq:matrixUncertaintyRel}), in canonical coordinates for two modes, are
\begin{eqnarray}
  \label{eq:35}
  \frac{1}{2}\left(C_1^2 + (U_x - U_y) \right) &=& \frac{\hbar^2}{4}, \\
  \frac{1}{2}\left(C_1^2 - (U_x - U_y) \right) &=& \frac{\hbar^2}{4}.
\end{eqnarray}
From these follow the two conditions
\begin{eqnarray}
  \label{eq:C12Gauss}
  C_1^2 &=& \frac{\hbar^2}{2}, \\ \label{eq:UxUy}
  U_x - U_y &=&0.
\end{eqnarray}
The first of these, because \(C_1^2 = \nu_+^2 + \nu_-^2\) is the total uncertainty, says that Gaussian states have minimum uncertainty. Moreover, because \(\nu_\pm\ge \hbar/2\), the equality (\ref{eq:C12Gauss}) implies \(\nu_+ = \nu_- = \hbar/2\) and consequently, also that \(C_2 = 0\) for a pure Gaussian state.

Using these constraints, the second condition (\ref{eq:UxUy}) can be expressed as
\begin{equation}
  \label{eq:38}
  4 p_\alpha p_\beta + \sqrt{(4p_\alpha)^2\left(p_\alpha^2-\frac{C_1^2}{2}\right)} \frac{\cos(\alpha)}{\sin(\beta)} = 0.
\end{equation}
This can be solved by either of
\begin{subequations}
  \begin{eqnarray}
    \label{eq:41}
    p_\alpha &=& 0 \\
    \label{eq:pbetaConstraint}
    p_\beta &=& \pm \sqrt{ p_\alpha^2-\frac{C_1^2}{2} } \frac{\cos(\alpha)}{\sin(\beta)}.
  \end{eqnarray}
\end{subequations}

The remaining off-diagonal equations in Eqs.~(\ref{eq:matrixUncertaintyRel}) are also solved by \(p_\alpha = 0\) in which case they do not present additional constraints. However, if we assume \(p_\alpha \ne 0\) and instead use the constraint (\ref{eq:pbetaConstraint}), the off-diagonal equations give the two additional constraints
\begin{equation}
  \label{eq:39}
  s_x p_{s_x} + s_y p_{s_y} = 0
\end{equation}
and
\begin{equation}
  \label{eq:40}
  p_\alpha = \frac{1}{2} \frac{\pm \sin(\alpha) \sqrt{\frac{\hbar^2}{4} (\sin^2(\alpha)-\cos^2(\beta)) + s_x^2p_{s_x}^2\sin^2(\beta)} - s_x p_{s_x} \sin(2\beta)}
  {\cos(2\alpha) + \cos(2\beta)}.
\end{equation}
The latter condition allows \(p_\alpha < C_1^2/2\) and, by (\ref{eq:pbetaConstraint}), would correspond to an imaginary solution for \(p_\beta\).

If we require all of the momenta solving Eqs.~(\ref{eq:matrixUncertaintyRel}) to be real, then we must rule out the solutions (\ref{eq:pbetaConstraint}), (\ref{eq:39}), and (\ref{eq:40}). Then the matrix equality (\ref{eq:matrixUncertaintyRel}) imposes only the three independent conditions
\begin{equation}
  \label{eq:appGauss1}
  C_1^2 = \frac{\hbar^2}{2},
\end{equation}
\begin{equation}
  \label{eq:appGauss2}
  C_2 = 0,
\end{equation}
\begin{equation}
  \label{eq:appGauss3}
  p_\alpha = 0.
\end{equation}

%\bibliographystyle{apsrev}
%\bibliography{references,../Bib/QuantGra,../Bib/Tunneling}

\end{document}